\documentclass[a4paper,11pt]{elsarticle}

\usepackage{setspace}
\usepackage{amssymb}
\usepackage{amsmath}
\usepackage{amsthm}
\usepackage[T1]{fontenc}
\usepackage{graphicx}
\usepackage[linesnumbered,ruled,vlined]{algorithm2e}
\usepackage[usenames]{color}
\usepackage{longtable}
\usepackage{booktabs}
\usepackage{url}
\usepackage{mathrsfs}
\usepackage{amsfonts}
\usepackage{enumerate}
\usepackage{multirow}
\usepackage{verbatim}
\usepackage[pagewise]{lineno}
\usepackage{makecell}
\usepackage{subfigure}
\usepackage{threeparttable}
\usepackage{diagbox}
\usepackage{lscape}
\usepackage{float}
\usepackage{caption}
\usepackage{natbib}
\usepackage{array}
\setcitestyle{authoryear,round}

\newtheorem{definition}{Definition}
\newtheorem{theorem}{Theorem}
\newtheorem{lemma}{Lemma}
\newtheorem{problem}{Problem}

\makeatletter
\def\ps@pprintTitle{%
   \let\@oddhead\@empty
   \let\@evenhead\@empty
   \let\@oddfoot\@empty
   \let\@evenfoot\@empty
}
\makeatother

\begin{document}
\onehalfspacing

\begin{frontmatter}



\title{A Branch-and-Bound Approach for Maximum Low-Diameter Dense Subgraph Problems}
\author[UESTC]{Yi Zhou\corref{cor}}
\cortext[cor]{Corresponding author} \ead{zhou.yi@uestc.edu.cn}

\author[UESTC]{Chunyu Luo}
\ead{Chunyu-Luo@std.uestc.edu.cn}

\author[PKU]{Zhengren Wang}
\ead{wzr@stu.pku.edu.cn}

\author[AIRS]{Zhang-Hua Fu} 
\ead{fuzhanghua@cuhk.edu.cn}

\affiliation[UESTC]{organization={University of Electronic Science and Technology 
 of China},
            city={Chengdu},
            country={China}}
\affiliation[PKU]{organization={Peking University},
            city={Beijing},
            country={China}}
\affiliation[AIRS]{organization={Shenzhen Institute of Artificial Intelligence and Robotics for Society, The Chinese University of Hong Kong, Shenzhen},
            city={Shenzhen},
            country={China}}

\begin{abstract}
A graph with $n$ vertices is an $f(\cdot)$-dense graph if it has at least $f(n)$ edges, $f(\cdot)$ being a well-defined function. 
The notion  $f(\cdot)$-dense graph encompasses various clique models like $\gamma$-quasi cliques, $k$-defective cliques, and dense cliques, arising in cohesive subgraph extraction applications. 
However, the $f(\cdot)$-dense graph may be disconnected or weakly connected.
To conquer this, we study the problem of finding the largest $f(\cdot)$-dense subgraph with a diameter of at most two in the paper. 
Specifically, we present a decomposition-based branch-and-bound algorithm to optimally solve this problem. 
The key feature of the algorithm is a decomposition framework that breaks the graph into $n$ smaller subgraphs, allowing independent searches in each subgraph. 
We also introduce decomposition strategies including degeneracy and two-hop degeneracy orderings, alongside a branch-and-bound algorithm with a novel sorting-based upper bound to solve each subproblem. 
Worst-case complexity for each component is provided. 
Empirical results on 139 real-world graphs under two $f(\cdot)$ functions show our algorithm outperforms the MIP solver and pure branch-and-bound, solving nearly twice as many instances optimally within one hour.
\end{abstract}




\end{frontmatter}



\section{Introduction}
\label{sec-introduction}

In a network, a \textit{cohesive subgraph}, or sometimes \textit{community}, is a group of closely connected members. 
Identifying large cohesive subgraphs is a crucial task in numerous real-world network applications, including community detection in complex networks~\citep{bedi2016community}, protein complex identification in biological networks~\citep{yu2006predicting}, and financial network statistical analysis~\citep{boginski2006mining}. 
The clique, which asks for an edge between every pair of vertices, naturally serves as a fundamental model for cohesive subgraphs. 
However, studies like ~\citet{balasundaram2011clique,pattillo2013clique} have shown that the clique model is often unsuitable in practical applications because the requirement for all possible connections to be present within the cohesive subgraph is overly strict, especially when noise and errors exist. 
Consequently, various relaxed clique models have been proposed in the literature \citep{abello2002massive,yu2006predicting,gao2022exact,chang2023efficient,wang2022listing,wang2023gap}.

To our knowledge, a notable number of relaxed clique models capture the number of edges in the induced graph.  
For example, a $\gamma$-quasi clique is a vertex set such that in its induced graph there are at least $\gamma \binom{n}{2}$ edges, where $n$ is the vertex number and $\gamma$ is a ratio in $(0,1]$ ~\citep{abello2002massive}. 
When $\gamma=1$, it is equal to the clique. 
Another example is the $s$-defective clique, which is a vertex set such that the induced graph has at least $\binom{n}{2}-s$ edges ~\citep{yu2006predicting,chen2021computing,luo2024faster}. 
Here, $s$ is a given non-negative integer. 
When $s=0$, the $s$-defective clique also degrades to a clique.
Other examples include \textit{average $s$-plexes} where the edges must be at least $\frac{n(n-s)}{2}$ and $s$ is a positive integer~\citep{veremyev2016exact}.
It is clear that the above clique models can be generalized by setting a general function as the lower bound on the number of edges, a general model initially referred to as the \textit{$f(\cdot)$-dense graph} \citep{veremyev2016exact}.
Specifically, an $f(\cdot)$-dense graph is defined as an $n$-vertex graph such that the number of edges is at least $f(n)$, with $f:\mathbb{Z}_{\ge0} \rightarrow \mathbb{R}$ being a function strictly smaller than $\binom{n}{2}$ for any $n\ge 0$. 



\begin{figure}[htb]
	\centering
        \scalebox{0.8}{
	\subfigure[$G_1$]{
		\begin{minipage}[b]{0.45\textwidth}
			\includegraphics[width=1\textwidth]{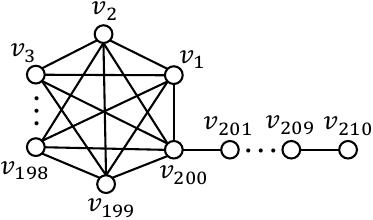} 
		\end{minipage}
		\label{subfig:G1}
	}
        \subfigure[$G_2$]{
            \begin{minipage}[b]{0.45\textwidth}
            \includegraphics[width=1\textwidth]{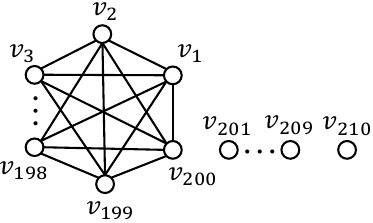}
            \end{minipage}
        \label{subfig:G2}
        }
        }
	\caption{$G_1[\{v_1,v_2,\dots,v_{200}\}]$ is a clique and $v_{200},v_{201},\dots,v_{210}$ forms a path between $v_{200}$ and $v_{210}$. $G_2[\{v_1,v_2,\dots,v_{200}\}]$ is a clique and $v_{201}, v_{202},\dots,v_{210}$ are all isolated vertices.}
	\label{fig:graph-exp}
\end{figure}

However, this simple generalization is inadequate for depicting cohesive groups, since for certain functions $f(\cdot)$, the graph with at least $f(n)$ edges may not be well-connected or might even be disconnected. 
For instance, in Fig. \ref{subfig:G1} and Fig. \ref{subfig:G2}, with $f(n)=0.9\binom{n}{2}$, both graphs are considered $f(\cdot)$-dense. However, in $G_1$, the path from vertex $v_1$ to $v_{210}$ requires 10 hops, and in $G_2$, there is no path between $v_1$ and $v_{210}$.
With respect to connectivity, these graphs are unlikely to be cohesive.
This issue has also been observed regarding the $\gamma$-quasi clique model. Hence, the connected constraint based on spanning tree was introduced in \citet{santos2024ensuring} . 

We introduce the \textit{low-diameter $f(\cdot)$-dense graph} and investigate the fundamental problem of finding the maximum low-diameter $f(\cdot)$-dense subgraph (M$f$DS) from a given graph.
In graph theory, the \textit{diameter} of a graph refers to the maximum length of the shortest path between any two vertices within the graph. 
A low-diameter $f(\cdot)$-dense graph refers to a connected graph where it has at least $f(n)$ edges and the diameter is no greater than two.
It is worth mentioning that a graph with a diameter limited to two was known as a $2$-club \citep{pattillo2013clique}, which is often studied alongside other structural constraints like vertex-connectivity \citep{veremyev2014finding,lu2022fault} in cohesive graph mining applications. 

\subsection{Related Works}

`Assuming that the $f(\cdot)$ can be computed in constant time, the M$f$DS problem is NP-hard, W[1]-hard, and hard to approximate in polynomial time. 
This conclusion follows directly from the fact that the maximum clique problem is a special case of M$f$DS~\citep{downey2012parameterized,haastad1999clique}.
Existing works related to the M$f$DS mainly focus on some special maximum $f(\cdot)$-dense graph, mostly overlooking the low-diameter constraint.
For example, the \textit{maximum $s$-defective clique} problem, which seeks the maximum $f(\cdot)$ subgraph given $f(n)=\binom{n}{2}-s$, where $s$ a fixed non-negative integer, was introduced in ~\citet{yu2006predicting}. 
The problem is NP-hard for any $s\ge 1$ ~\citep{yannakakis1978node}. 
\citet{luo2024faster} showed that the maximum $s$-defective clique with low-diameter constraint can be solved in $O(n^c1.2^n)$ where $n^c$ and $1.2^n$ represent polynomial and exponential factors of $n$, respectively.
Another example, the \textit{maximum $\gamma$-quasi clique} problem, which is the maximum $f(\cdot)$-dense subgraph problem given $f(n)=\gamma \binom{n}{2}$ and $\gamma$ being a real number within $(0,1]$, was introduced in ~\citet{abello2002massive}. 
This problem also remains NP-hard for any fixed $\gamma \in (0,1]$ \citep{pattillo2013maximum}.
As a matter of fact, if $f(n)=\Theta(n^{1+\epsilon})$, where $\epsilon \in (0,1)$  or $f(n)=n+\Omega(n^{\epsilon})$, where $\epsilon \in (0,1)$, the problem of finding a maximum $f(\cdot)$-dense graph  is always NP-hard~\citep{asahiro2002complexity,holzapfel2003complexity}.  
Nevertheless, there are some polynomial solvable cases, such as when $f(n)=n+c$ and $c$ is a constant \citep{asahiro2002complexity}.


Although the hardness results seem to be discouraging, empirically solving the M$f$DS problem optimally on real-world graphs seems feasible.
One of the most common practical approaches is \textit{mixed integer programming} (MIP).
To the best of our knowledge, specific MIP formulations for the maximum $s$-defective clique problem and the maximum $\gamma$-quasi clique were proposed by ~\citet{sherali2002airspace} and ~\citet{pattillo2013maximum}, respectively.
Later, ~\citet{veremyev2016exact} investigated different MIP formulations for the general maximum $f(\cdot)$-dense subgraph problem. 
In ~\citet{veremyev2016exact}, empirical experiments showed that an MIP model called \textit{GF3} outperforms other models across various configurations. 
On the other hand, MIP-based solutions for the maximum 2-club problem, which aims to find the largest 2-club from the given graph, were studied in \citet{pajouh20162,lu2022fault,moradi2018finding,almeida2014analytical}.
However, to date, there is no available MIP formulation for solving the maximum low-diameter $f(\cdot)$-dense subgraph problem.
Furthermore, the efficiency of MIPs remains a bottleneck, particularly when the input real-world graph becomes very large.

On the other hand, combinatorial search methods such as branch-and-bound  are another popular tool for solving large-scale instances exactly. 
For example, the branch-and-bound algorithms for maximum $s$-defective clique problem ~\citep{chen2021computing, gao2022exact, zheng2023kd, chang2023efficient}, maximum $\gamma$-quasi clique problem ~\citep{mahdavi2014branch, ribeiro2019exact} and even the maximum 2-club problem  \citep{hartung2012parameterized,komusiewicz2019exact,carvalho2011upper} have been shown to largely outperform their MIP counterparts.
When the guarantee of optimality can be sacrificed, heuristic combinatorial algorithms, such as local search, are often used for solving these subgraph extraction problems.
For example, local search  ~\citep{chen2021nuqclq} and meta-heuristic algorithms  ~\citep{zhou2020opposition,pinto2018biased}  approximate the size of maximum quasi-cliques. 
Interestingly, even for graphs with millions of vertices, branch-and-bound algorithms have demonstrated the ability to find optimal solutions within hours or even minutes \citep{luo2024faster,wang2023gap,walteros2020maximum}. 
This suggests that employing branch-and-bound algorithms to derive optimal solutions for large-scale, real-world M$f$DS problems remains a feasible approach.

\subsection{Our Contribution}
Our contributions in this work are summarized as follows.

\begin{itemize}
\item \textit{Formalization of Maximum Low-diameter \( f(\cdot) \)-Dense Subgraph Problem}:  
   We introduce the \( f(\cdot) \)-dense graph, defined as a  graph with at least \( f(n) \) edges, where \( f(\cdot) \) is a well-defined function and \( n \) the number of vertices. 
   This model encompasses various cohesive graph models and is  discussed in terms of the hereditary property. We define the maximum low-diameter \( f(\cdot) \)-dense subgraph problem, which seeks the largest \( f(\cdot) \)-dense subgraph problem with a diameter bounded by two, ensuring connectivity. A mixed integer linear programming model for solving this problem is also provided.
\item \textit{Algorithmic Framework}:
   We propose an efficient framework to solve the M\( f \)DS problem optimally. The framework includes a graph decomposition pre-processing step, which involves breaking the input graph into smaller subgraphs, enabling independent solutions for each subgraph. To ensure the subgraphs are small, we introduce \textit{degeneracy ordering} and \textit{two-hop degeneracy ordering}. Additionally, we present the branch-and-bound algorithm with reduction strategies and an efficient sorting-based upper bounding method to improve performance. Worst-case complexity analyses for each component and the overall algorithm are provided.

\item \textit{Empirical Evaluation}:
   We conduct experiments on 139 real-world graphs with two \( f(\cdot) \) functions. The results show that our graph decomposition combined with branch-and-bound outperforms methods like the MIP solver and pure branch-and-bound. With decomposition, we solve about twice as many instances within one hour. We also show that the sorting-based upper bounding significantly speeds up the algorithm by reducing the search space. 
\end{itemize}

\section{Preliminary}\label{preliminary}
\subsection{Basic Notation}
Let \( G = (V, E) \) be an undirected simple graph, where \( V \) is the set of vertices and \( E \) is the set of edges. 
For any set \( S \), we use \( |S| \) to represent the cardinality of \( S \), which is the number of elements in the set.

For a vertex $u\in V$,  $N_G(u)$ denotes the set of \textit{neighbor vertices} of $u$ in graph $G$.
Therefore, $|N_G(u)|$ denotes the \textit{degree} of $u$ in $G$, which is the number of neighboring vertices of $u$.
We use $\Delta_G$ to denote the maximum degree over all vertices in $G$, i.e., the largest value of $|N_G(u)|$ over all vertices $u\in V$.
We denote $dist_G(u,v)$ as the distance between two vertices $u$ and $v$ in $G$, defined as the length of the shortest path from $u$ to $v$.
If $dist_G(u,v)\le k$, then we say $v$ is a $k$-hop neighbor of $u$.
Moreover, the set of \textit{two-hop neighbors} of $u$ in $G$ consists of vertices that can be reached from vertex $u$ by traversing at most two edges in $G$, denoted by $N_G^{(2)}(u)$. 
Note that $N_G(u)\subseteq N^{(2)}_G(u)$, since all direct neighbors of $u$ are also two-hop neighbors.


Given a vertex subset $S\subseteq V$, $G[S]$ denotes the \textit{subgraph induced by $S$}. Specifically,  $E(S)$ denotes the edge set in $G[S]$, which includes all edges spanned by $S$.
We use $\overline{G}=(V,\overline{E})$ to denote the \textit{complement graph} of $G=(V,E)$. The complement graph $\overline{G}$ has the same vertex set $V$, but an edge $(u,v)$ exists in $\overline{E}$ if and only if $(u,v)\notin E$.
Likewise, for a subset $S\subseteq V$, we use $\overline{G[S]}$ to denote the subgraph induced by $\overline{G}$ and $\overline{E(S)}$ to denote the edge set of $\overline{G[S]}$, which includes all edges spanned by  $S$ in the complement graph $\overline{G}$.

Given a graph \( G = (V, E) \), the \textit{diameter} of \( G \), denoted by \( diam(G) \), is the maximum length of the shortest paths between any pair of vertices in \( G \). A \textit{clique} in \( G \) is a subset \( S \subseteq V \) such that any two distinct vertices in \( S \) are adjacent in \( G \). In contrast, an \textit{independent set} in \( G \) is a subset \( S \subseteq V \) such that no two distinct vertices in \( S \) are adjacent in \( G \).


\section{The Maximum $f(\cdot)$-dense Subgraph Problem}\label{sec-problem-form}

In this section, we formalize the concept of $f(\cdot)$-dense graph and define the maximum low-diameter $f(\cdot)$-dense subgraph problem. Furthermore, we formulate the problem as a mixed integer programming model.

\subsection{$f(\cdot)$-Dense Graph}

We formally define the problem from the \textit{density function}.

\begin{definition}[density function]\label{def:dense-graph function}
        A function $f:\mathbb{Z}_{\ge0} \rightarrow \mathbb{R}$ defined on the non-negative integers is called a density function if, for all $i \ge 0$, $f(i) \le \binom{i}{2}$. 
\end{definition}

In the following, we use a density function $f(i)$ to bound the minimum number of edges in a graph of $i$ vertices. 
The $f(i)$ is at most $\binom{i}{2}$ as there are at most $\binom{i}{2}$ edges in the graph.


\begin{definition}[$f(\cdot)$-dense graph] \label{def:dense-sub}
        Given a density function $f:\mathbb{Z}_{\ge 0}\rightarrow \mathbb{R}$ and a graph $G=(V,E)$ with $n$ vertices and $m$ edges, if $m \ge f(n)$, then $G$ is called an $f(\cdot)$-dense graph.
\end{definition}
 
Clearly, a clique is a $f(\cdot)$-dense graph for any density function $f(\cdot)$ because $\binom{i}{2}\ge f(i)$ for all $i\ge 0$.
The $f(\cdot)$-dense graph also generalizes many well-known relaxed clique models, as discussed in the introduction.

\subsubsection{Hereditary Property of $f(\cdot)$-Dense Graph}
A graph property $\Pi$ is said to be \textit{hereditary} if, whenever a graph $G$ has property $\Pi$, any induced subgraphs of $G$ also has property $\Pi$.
The hereditary property plays an important role in the design of subgraph extraction algorithms.

\begin{definition}[hereditary-induced  density function]\label{def:h-dense-f}     
    Given a density function $f:\mathbb{Z}_{\ge 0}\rightarrow \mathbb{R}$ and an $f(\cdot)$-dense graph $G$, if every induced subgraph of $G$ is also an $f(\cdot)$-dense graph, then  $f(\cdot)$ is a hereditary-induced density function, or simply a hereditary-induced function. 
\end{definition}

    \begin{lemma}\label{ppt:H}
         Given a density function $f:\mathbb{Z}_{\ge0}\rightarrow \mathbb{R}$, define $g_f(i)=\binom{i}{2}-f(i)$, for all $i\in \mathbb{Z}_{\ge0}$.
         If $g_f(\cdot)$ is a monotonically non-increasing, i.e., $g_f(i)\ge g_f(i+1)$, for all $i\in \mathbb{Z}_{\ge 0}$, then $f(\cdot)$ is a hereditary-induced function.
    \end{lemma}
    The proof of this lemma, along with the subsequent lemmas and theorems, is deferred to  \ref{sec_missing_proof}.

    \begin{lemma}\label{ppt:non-H}
        Given a density function $f:\mathbb{Z}_{\ge0}\rightarrow \mathbb{R}$, suppose the following conditions hold: 
        \begin{enumerate}
            \item $g_f(\cdot)$ is positive, i.e. there exists some $n_0\ge0$ such that $f(n_0)>0$,
            \item $g_f(\cdot)$ is monotonically non-decreasing , i.e. for all $i \in \mathbb{Z}_{\ge0}, g_f(i)\le g_f(i+1)$, 
            \item $g_f(\cdot)$ is unbounded, i.e. for any constant $c\in \mathbb{R}_{\ge0}$, there exits some $n\in \mathbb{Z}_{\ge0}$ such that $g_f(n)\ge c$.
        \end{enumerate}
        Then $f(\cdot)$ is not a hereditary-induced function. 
    \end{lemma}

    We show some examples in the following.
    \begin{itemize}
        \item Define $f(i)=\gamma \frac{i(i-1)}{2}$, where $\gamma\in (0,1)$. Clearly $g_{f}(\cdot)$ is positive, monotonically non-decreasing and unbounded. By Lemma \ref{ppt:non-H}, $f(\cdot)$ is not a hereditary-induced function. (This indicates that the $\gamma$-quasi-clique is not hereditary.)
        \item Define $f(i) = \frac{i(i-1)}{2} - s$, where $s \in \mathbb{Z}_{\ge 0}$. Since $g_f(\cdot)$ is monotonically non-increasing, by Lemma \ref{ppt:H}, $f(\cdot)$ is a hereditary-induced function. (This indicates that the $s$-defective clique is hereditary.)
        \item Define  $f(i) = \frac{i(i-1)}{2} - \log(i+1)$. Here, $g_f(\cdot)$ is positive, monotonically non-decreasing, and unbounded. By Lemma \ref{ppt:non-H}, $f(\cdot)$ is not a hereditary-induced function. (This indicates that the graph property with this specific function is not hereditary.)
    \end{itemize}


\subsection{The Maximum Low-Diameter $f(\cdot)$-Dense Subgraph Problem}

Now, we introduce the maximum low-diameter $f(\cdot)$-dense subgraph problem (M$f$DS).

\begin{problem}[M$f$DS]
Given a graph $G=(V,E)$ and a density function $f(\cdot)$, the maximum low-diameter $f(\cdot)$-dense subgraph problem asks for the largest subset of vertices $S$ such that it induces an $f(\cdot)$-dense graph and $diam(G[S])\le 2$.
\end{problem}
 If the context is clear, we abuse the notation a bit by saying that $S\subseteq V$ is a low-diameter $f(\cdot)$-dense subgraph.
The M$f$DS problem can be further expressed as the following.
\[
\omega_f(G)=\max \{|S|: S\subseteq V, |E(S)|\ge f(|S|) \text{ and } diam(G[S])\le 2\}
\]

As stated in the introduction, the purpose of imposing additional diameter constraints is to exclude certain poorly connected or disconnected graphs, such as $G_1$ and $G_2$ shown in Fig. \ref{fig:graph-exp}.

Given an $f(\cdot)$-dense subgraph in $G$, if $f(\cdot)$ is not hereditary-induced, then it maybe even NP-hard to decide if the graph is maximal \citep{pattillo2013clique}.
If $f(\cdot)$ is a hereditary-induced function, given an $f(\cdot)$-dense graph $S$, it is easy to check if any vertex not in $S$ can form a larger $f(\cdot)$-dense graph with $S$. 
In other words, whether a given vertex set $S$ is  maximal can be checked in polynomial time in this case.
Unfortunately, the low-diameter $f(\cdot)$-dense graph is not hereditary because the induced subgraph of a diameter-2 bounded graph may not be a  diameter-2 bounded subgraph. 
However, if the $f(\cdot)$-density function is hereditary-induced, the algorithm in the remainder of the paper still uses some extra pruning (reduction) rules.

\subsection{A Mixed Integer Programm of M$f$DS}
\label{section_mip}
As mentioned, \citet{veremyev2016exact} proposed multiple mixed integer programs (MIP) for the $f(\cdot)$-dense subgraph problem (without low-diameter constraint) and identified one model, \textit{GF3}, as the most efficient. 
We extend this model by further encoding the diameter constraint, resulting in a new model,  MIP-$f$D, specifically designed to solve the M$f$DS problem.

\begin{align}
    \text{(MIP-$f$D)} \max  & \quad \sum_{i\in V}{x_i}  \nonumber\\
    \text{s.t.} &  \sum_{(i,j)\in E}y_{ij}\geq \sum_{k=\omega_l}^{\omega_u}f(k)z_k \label{f-dense-constraint}\\
    & y_{ij}\leq x_i, y_{ij}\leq x_j, \forall (i,j)\in E   \label{constraint_conflict} \\
    & \sum_{i\in V}x_i=\sum_{k=\omega_l}^{\omega_u}kz_k, \sum_{k=\omega_l}^{\omega_u}z_k=1 \label{constraint_atmost1zk}\\
    & x_i+x_j\leq 1  \text{\ \ \ \ \ \ \ \ \ } \forall i,j\in V, dist_G(i,j)>2  \label{constaint_dist2a}\\
    & x_i+x_j-\sum_{r\in N(v_i)\cap N(v_j)}x_r\leq 1 \text{\ \ } \forall i,j\in V,\ dist_G(i,j)= 2  \label{constaint_dist2b}\\
    & x_i\in \{0,1\},  \forall i \in V\\
    & y_{ij}\geq 0, \forall (i, j)\in E, z_k\geq 0,\ \forall k \in \{\omega_l,\dots,\omega_u\}
\end{align}

In MIP-$f$D, the binary variable $x_i$ represents whether the vertex $i$ is chosen in the solution, and $y_{ij}$ represents whether there exists an edge between vertex $i$ and $j$ in $G$. 
The binary variable $z_k$ indicates whether $\omega_k$ ($k\in \{\omega_l,\dots,\omega_u\}$) is equal to the size of the optimal size.
$\omega_l$ and $\omega_u$ are two given parameters, representing the lower and upper bounds of optimal size, respectively. 
The Inequality \ref{f-dense-constraint} indicates that the edge number of the solution subgraph is at least $f(\cdot)$.
The Inequality \ref{constraint_conflict} stipulates that the edges must be in the induced graph of a solution (vertex) set.
The Inequality \ref{constraint_atmost1zk} asks that only one $z_k$ can be 1 for all $k$ in the range $\{\omega_l,\omega_u\}$. 
In brief, Inequalities \ref{f-dense-constraint}-\ref{constraint_atmost1zk} encode that the final solution set induced a subgraph that has at least $f(\cdot)$ edges.
For any two vertices $i$ and $j$ in $V$, $dist_G(i,j)$ in Inequality \ref{constaint_dist2a} denotes the length of the shortest path between $i$ and $j$ in $G$.
The Inequalities \ref{constaint_dist2a} and \ref{constaint_dist2b} require that the distance between any two vertices in the induced subgraph is bounded by 2.
Note that in the model, $k$ is a constant, so MIP-$f$D is a mixed integer linear program.

To build the model, we need to establish a lower bound $\omega_l$ and upper bound $\omega_u$ for the size of the maximum low-diameter $f(\cdot)$-dense subgraph. 
While the trivial bounds $\omega_l=1$ and $\omega_u=|V|$ would suffice, the performance of the MIP solver can be significantly improved with tighter bounds.
To achieve these tighter bounds, we use the efficient heuristic described in Section \ref{subsec:HeuAlg} to obtain a feasible solution. The size of this solution provides a good lower bound, which we set as $\omega_l$. Then, we apply the bounding algorithm outlined in Section \ref{subsec:FastUB} to estimate an upper bound $\omega_u$, which is better than the trivial upper bound $|V|$.

\section{A Combinatorial Algorithm of Solving MfDS}\label{sec:BB-Framework}

\subsection{The Decomposition-based Framework}\label{subsec:AlgFramework}

\begin{algorithm}[htb]
	\footnotesize
	\caption{The overall decomposition-based framework for solving the M$f$DS problem.}\label{Framework}
	\KwIn{A graph $G=(V,E)$ and an oracle to $f(\cdot)$}
	\KwOut{The maximum $f(\cdot)$-dense subgraph $S$.}
        OptM$f$DS($G=(V,E), f(\cdot)$)\\
        \Begin{
            $S\gets $ an initial heuristic solution \tcc{Section \ref{subsec:HeuAlg}}
            $\pi\gets $  an order the vertices in $G$ by a ordering algorithm.             Suppose  $\pi=(v_1,\dots,v_n)$ w.l.o.g. \tcc{Section \ref{subsec:ordering}}
            \For{$v_i$ from $v_1$ to $v_n$}{       
                $G_i=(V_i,E_i)$ be the graph induced by $\{v_i\} \cup \left\{N_G^{(2)}(v_i) \cap  \{v_{i+1},\dots,v_n\} \right\} $ \\
                $S_i^*\gets$ ExactSearch$(G_i, f(\cdot), \{v_i\}, N_G^{(2)}(v_i),lb)$ \tcc{Section \ref{subsec:BBSearch}}                
            }
    	\Return{$\max (|S_1^*|,\dots,|S_n^*|)$}
        }
\end{algorithm}

In this section, we present an algorithmic framework for solving the M\( f \)DS problem, as outlined in Algorithm \ref{Framework}. The input to the framework consists of a graph \( G = (V, E) \) and an oracle that returns the value of the function \( f(\cdot) \) for any given \( i \in \{0, \ldots, |V|\} \).

The framework begins by determining a heuristic solution \( S \), where the size of \( S \) provides a lower bound for future searches. 
The lower bound $|S|$ can be also used to prune unfruitful searches. 
A key feature of the framework is its graph decomposition strategy, which is based on the following simple yet crucial observation.

\begin{lemma}\label{lemma:decompose}
    Let $G=(V,E)$ be a graph with $n=|V|$. Denote $\pi=<v_1,\dots,v_n >$ as an arbitrary order of $V$. 
    For each vertex $v_i$ where $i=1,\dots,n$, let $V_i$ be $\{v_i\}\cup \left\{ N_G^{(2)}(v_i) \cap  \{v_{i+1},\dots,v_n\}\right\}$. 
    Suppose that $S^*_i$ is a maximum low-diameter $f(\cdot)$-dense graph in $G[V_i]$ that must contain $v_i$. Then, we have $\omega_f(G) = \max (|S_1^*|,\ldots,|S_n^*|)$.
\end{lemma}

Note that the vertex set $V_i$ consists of the vertex $v_i$ itself, along with the vertices in $N_G^{(2)}(v_i)$ that are ranked after $v_i$ in the order $\pi$.

In Alg. \ref{Framework}, for each vertex $v_i$ in $\pi$, we construct a subgraph induced by $V_i$, i.e. $G_i=(V_i,E_i)$, and then apply an exact algorithm, referred to as ExactSearch, to find the largest $f(\cdot)$-dense subgraph that includes $v_i$ from $G_i$. According to Lemma \ref{lemma:decompose}, the optimal $f(\cdot)$-dense graph of the entire graph $G$ is the largest subgraph obtained from this process across all $v_i$.

Assume that the oracle computes \( f(i) \) for any given integer \( i \) in constant time, and the ExactSearch algorithm checks all subsets of vertices in $G_i$ in the worst case. 
Define \( \alpha_G = \max(|V_1|, \ldots, |V_n|) \). Then, the total running time of the algorithm is
\[
O(T_{heur}(G) + T_{order}(G) + poly(\alpha_G) 2^{\alpha_G}),
\]
where $poly(\alpha_G)$ indicates a polynomial factor of $\alpha_G$, and \( T_{heur}(G) \), \( T_{order}(G) \) and $poly(\alpha_G) 2^{\alpha_G}$ represent the time required to find a heuristic solution, order the vertices of \( G \), and run ExactSearch, respectively. 

From a complexity perspective, an efficient polynomial-time heuristic algorithm is easy to obtain, as we suggest in Section \ref{subsec:HeuAlg}. Therefore, the focus is on making the sorting algorithms efficient, and, more importantly, on minimizing \( \alpha_G \), due to the exponential part related to \( \alpha_G \). (Indeed, studies in the maximum clique problem \citep{wang2022listing} also suggest that smaller subgraphs improve memory access performance in large graphs.) 
In Section \ref{subsec:ordering}, we introduce two polynomial-time ordering algorithms that ensure \( \alpha_G \) remains sufficiently small in practice.

\subsection{A Efficient Heuristic Algorithm}\label{subsec:HeuAlg}

Given a graph $G=(V,E)$ and an oracle to the density function $f(\cdot)$, we introduce an efficient heuristic, as shown in Alg. \ref{alg:Heualg}, that returns a low-diameter $f$-dense subgraph $S$. 
The algorithm relies on a \textit{degeneracy ordering} of graph $G$.

\begin{definition} [degeneracy ordering]
A $k$-degeneracy ordering of a graph \(G\) is an ordering of the vertices of \(G\), say $\langle v_1, v_2, \ldots, v_n \rangle$, such that for each vertex \(v_i\), its neighbors in the subgraph induced by \(\{v_i, v_{i+1}, \ldots, v_n\}\) are at most \(k\). 
A degeneracy ordering of $G$ is a $k$-degeneracy ordering such that $k$ is minimized, and the \textit{degeneracy of \(G\)} is such a minimum value $k$, denoted by $d_{G}$.
\end{definition}

A well-known \textit{peel} algorithm determines a degeneracy ordering in linear time  \citep{matula1983smallest}. 
Given a graph $G$, the peel algorithm repeatedly removes the vertex with the lowest degree in the graph, and this process ends when all the vertices are removed. 
The peel then determines the degeneracy order of a vertex at its removal, i.e., the $i$th removed vertex ranks $i$th in the final ordering. 
Using a linear-heap data structure, the peel algorithm has running time $O(|V|+|E|)$.

Given a degeneracy order $\langle v_1,\dots,v_n\rangle$, define the subgraph induced by  $\{v_i\} \cup \{N_G(v_i)\cap \{v_{i+1},\dots,v_n\}\}$ as $G_i=(S_i, E_i)$.
Clearly, the diameter of $G_i$ is at most two because all vertices in $S_i$ except $v_i$ are neighbors of $v_i$.
However, $G_i$ does not necessarily satisfy $|E_i|\ge f(|S_i|)$, so it is not necessarily an $f(\cdot)$-dense graph.
Observing this, for each $G_i$, our heuristic algorithm repeatedly deletes the vertex with the smallest degree in $G_i$ again, until the remaining graph satisfies that $|E_i|\geq f(|S_i|)$. 
 
Note that the deletion does not remove $v_i$ because the degree of $v_i$ is the largest in $G_i$. (If all vertices in $G_i$ have the same degree as $v_i$, then $G_i$ is a clique.)
Therefore, for each $G_i$, the deletion procedure finishes with a low-diameter $f(\cdot)$-dense subgraph. 
The largest subgraph found by this deletion procedure for all $G_i$s is returned as the solution.


\begin{algorithm}[htb]
	\footnotesize
	\caption{A heuristic algorithm for the M$f$DS problem.}\label{alg:Heualg}
	\KwIn{A graph $G=(V,E)$ and an oracle to $f(\cdot)$}
	\KwOut{A low-diameter subgrapgh $H^*$ }
    InitHeuristic($G=(V,E),f(\cdot)$)
    \Begin{
    $S^*\gets \emptyset$\\
    Compute a degeneracy ordering for the vertices of $G$ by peeling, denote the ordering as $(v_1,\dots,v_n)$\\
    \For{$v_i$ from $v_1$ to $v_n$}{
        $S_i\gets \{v_i\} \cup \{N_G(v_i)\cap \{v_{i+1},\dots,v_n\}\}$\\
        $G_{i}=(S_i,E_i)\gets$ the subgarph induced by $S$ in $G$. \\
        \While{$|E_i|< f(|S_i|)$}{
            $v\gets$ the vertex with the minimum degree in $G_i$\\
            $S_i \gets S_i\setminus\{v\}$
        }
        \If{$|S_i|>|S^*|$}{
            $S^*\gets S_i$
        }
    }
	\Return{$S^*$}
    }
\end{algorithm}

In the following, we give a time complexity analysis of the heuristic algorithm. 
The degeneracy ordering time is $O(|V|+|E|)$. 
For every vertex $v_i$, $|S_i|$ is at most $d_G+1$ and then $|E_i|$ is bounded by $O(d_G^2)$. 
Therefore, the time to build subgraph $G_i$ is bounded by $O(d_G^2)$.
The procedure in lines 7-9, which keeps deleting the smallest vertex from $G_i$, works the same way as the peel algorithm. 
Therefore, the time complexity of this procedure for any $G_i$ is $O(|V_i|+|E_i|)$, or simply $O(d_G^2)$.
Combining these elements, the overall complexity is $T_{heur}=O(|E|+|V|d_G^2)$.

In fact, the values of $d_G$ in real instances are quite smaller than $|V|$. For example, for the real-world graph \textit{ca-dblp-2010}, $d_G=74$, in contrast to $|V|=226413$.

\subsection{The Ordering of Vertices}
\label{subsec:ordering}
In this section, we introduce two specific orderings used for the graph decomposition.
\subsubsection{Degeneracy Ordering}
\label{subsubsec:degeneracyordering}

In the above heuristic algorithm, we propose the degeneracy order.
Because  $d_G$ is much smaller than $|V|$, it is reasonable to use the degeneracy order again for the ordering of the vertices in line 4 in Algorithm \ref{Framework}.

\begin{theorem}
If degeneracy ordering is applied as the vertex ordering in Line 4 in Algorithm \ref{Framework}, then $T_{order}= O(|V|+|E|)$ and $\alpha_{G}\le \min (d_G\Delta_G, |V|)$.
\end{theorem}


\subsubsection{Two-hop Degeneracy Ordering}\label{subsubsubsec:weak-degen}

Another idea of ordering of the vertices is called \textit{two-hop degeneracy ordering}.  


\begin{definition}[two-hop degeneracy ordering]\label{2hop-degen-order}
A $k$-two-hop degeneracy ordering of a graph \(G\) is an ordering of the vertices of \(G\), say \(v_1, v_2, \ldots, v_n\), such that for each vertex \(v_i\), its two-hop neighbors in the subgraph induced by \(\{v_i, v_{i+1}, \ldots, v_n\}\) is at most \(k\). 
Likewise, a two-hop degeneracy ordering of $G$ is a $k$ two-hop degeneracy ordering such that $k$ is minimized, and the \textit{two-hop degeneracy of \(G\)} is this minimum value $k$, denoted by $td_G$.
\end{definition}

\begin{algorithm}[htb]
	\footnotesize
	\caption{The algorithm for computing the two-hop degeneracy order of a graph.}\label{alg_weak_degeneracy}
	\KwIn{A graph $G=(V,E)$}
	\KwOut{An ordering of $V$ and the two-hop degeneracy ordering of $G$ .}
    TwoHopDegeneracy($G=(V,E)$)\\
    \Begin{
    \For{each $v\in V$}{
        Run a two-depth breadth first search (BFS) starting from $v$ to compute $|N^{(2)}_G(v)|$\\
        Let $tdeg(v)\gets |N^{(2)}_G(v)|$
    }
    Let $td_G \gets 0,\ Ord \gets \emptyset$\\
    $\pi \gets $ an empty sequence\\
    \For{$i\gets 1$ to $n$}{
        $u\gets \arg\min_{v\in V\setminus Ord}tdeg(v)$\\
        append $u$ to the tail of $\pi$\\
        \If {$tdeg(v)\ge td_G$}{
            $td_G\gets tdeg(v)$
        }
        \For{each $v$ in $N^{(2)}_G(u)\setminus N_G(u)$}{
            $tdeg(v)\gets tdeg(v)-1$
        }
        \For{each $v$ in $N_G(u)$}{
            Run two-depth BFS starting from $v$ to compute $|N^{(2)}_G(v)|$\\
        }
    }
    \Return{$\pi$}
    }
\end{algorithm}

In Alg. \ref{alg_weak_degeneracy} we show an algorithm to compute the two-hop degeneracy order of a given graph.
In general, the two-hop degeneracy ordering is similar to that of peel. 
We keep track of the $|N^{(2)}_G(v)|$ value of each vertex, and then iteratively remove a vertex with the smallest $|N^{(2)}_G(v)|$ value. The two-hop degeneracy order is the sequence of removing these vertices.

The running time of removing a vertex $u$ and updating all $tdeg(v)$ is $O(td_G + m\min{(\Delta_G, td_G)})$, 
where the first factor is from $|N^{(2)}_G(u)|\le td_G$, and the second factor is from $|N_G(u)|\le \min {(\Delta_G, td_G)}$. 
Therefore, the worst-case running time of Alg. \ref{alg_weak_degeneracy} is 
\[
O(|V||E|+\sum_{i=1}^{|V|}{(td_G + |E|\min {(\Delta_G, td_G)})})=O(|V||E|\min{(td_G,\Delta_G)})
\]



\begin{theorem}
If the two-hop degeneracy ordering is used as the vertex ordering  in Line 4 in Algorithm \ref{Framework}, then $T_{order}= O(|V||E|\min{(td_G,\Delta_G)})$ and $\alpha_{G}\le td_G$.
\end{theorem}

We observed that in large sparse graphs, $td_G$ is usually much smaller than $\Delta_G d_G$ and $|V|$. 
For example, for the real-world graph \textit{ca-dblp-2010}, the value $td_G$ is $238$, which is much smaller than $|V|=226413$ or $\Delta_G d_G=17612$.
Note that the concept of two-hop degeneracy ordering was initially proposed in ~\citet{wunsche2021mind} (using the name \textit{weakly degeneracy ordering}). 
The empirical performance of this ordering for finding maximum \textit{$k$-plex} and \textit{$k$-club} was reported in ~\citet{figiel_et_al:LIPIcs.ESA.2023.47}. 

\subsection{The Branch-and-Bound Algorithm} \label{subsec:BBSearch}

\begin{algorithm}[htb]
	\footnotesize
	\caption{ExactSearch (An exact branch-and-bound algorithm.)}\label{alg-branch-bound}
	\KwIn{Graph $G=(V,E)$, an oracle to $f(\cdot)$ and a lower bound solution $lb\ge 0$}
	\KwOut{The maximum low-diameter $f(\cdot)$-dense subgraph $S$.}
         BranchBound($G=(V,E),f(\cdot), S,C,lb$)\\
         \Begin{
         \If{$|S|>lb$ and $diam(G[S])\leq 2$ and $|E(S)|\geq f(|S|)$}{
            $lb\gets |S|$
         }
         \If{$f(\cdot)$ is hereditary-induced and $|E(S)|< f(|S|)$}{ 
            \Return{} \tcc{Based on case 1 in Lemma \ref{lemma_hereditary_reduction}}
         }
         $UB\gets \text{SortBound($G,f(\cdot), S,C$)}$\\
         \If{$UB> lb$}{
            \If{$f(\cdot)$ is hereditary-induced}{
                Delete the vertices and edges by cases 2 in Lemma \ref{lemma_hereditary_reduction}
             }
            Randomly pick a vertex $u$ from $C$\\
            \If{$|N_G(u)|\geq |V|-2$, $|E(S\cup \{u\})|\geq f(|S|+1)$ and $diam(G[S\cup \{u\}])\leq 2$}{
                BranchBound($G=(V,E),f(\cdot), S\cup\{u\},C\setminus\{u\},lb$) \tcc{Based on Lemma \ref{lemma:high-deg-reduce}}
            }
            \Else{
                BranchBound($G=(V,E),f(\cdot), S\cup\{u\},C\setminus\{u\},lb$)\\
                BranchBound($G=(V,E),f(\cdot), S,C\setminus\{u\},lb$)\\
            }
         }

        } 
\end{algorithm}
       
In this section, we show a branch-and-bound algorithm in Alg. \ref{alg-branch-bound} for exactly solving the M$f$DS.
It is well-known that the branch-and-bound is essentially a tree-search algorithm. 
Each recursive call to the algorithm BranchBound($G,S,C,lb$) solves a \textit{restricted problem}: 
Given an input graph $G=(V,E)$, an oracle to density function $f(\cdot)$,  a growing set $S\subseteq V$,  a candidate set $C\subseteq V$, and an integer $lb$, return the largest maximum low-diameter $f(\cdot)$-dense graph $G[S^*]$ such that $S\subseteq S^*\subseteq S\cup C$ and $|S|\ge lb$; if such a low-diameter $f(\cdot)$-dense graph does not exist, then just return \textit{none}.
This restricted problem clearly generalizes the original problem because we can solve the original problem by setting $S=\emptyset$, $C=V$ and $lb=0$. 

For convenience purposes, we maintain the lower bound $lb$ and the best-known vertex set $S^*$ as globally accessible variables in the branch-and-bound.  
Note that the values of $lb$ and $S^*$ set are initialized by the heuristic algorithm in Section \ref{subsec:HeuAlg}, and updated in the recursive call to BranchBound.
Following the conventions, we call each recursive call to BranchBound as a node of the search  tree.
Then, in each node of the tree,  if $G[S]$ is a low-diameter $f(\cdot)$-dense subgraph and $|S| \ge lb$, we update $lb$  and  $S^*$. 
This is done in line 4 in the algorithm.
After updating $lb$, the BranchBound continues by either halting the search  as described in lines 5-6 (see Section \ref{subsec:reducing_vertices} for details) , or estimating an upper bound  by the so-called \textit{SortBound} algorithms (see Section \ref{subsec:FastUB} for details).
If the search goes on after checking the bound, then some unfruitful vertices of $C$ can be reduced, as described in lines 9-10, and a vertex $u$ is randomly chosen from $C$, as described in line 11. 
Lastly, we either reduce the search by moving $u$ into the solution $S$, as described in lines 12-13 (see Section \ref{subsec:reducing_vertices} for details), or branch the search into two subnodes, as described in lines 15-16.

\subsection{Reducing Unfruitful Vertex}
\label{subsec:reducing_vertices}

The correctness of reductions relies on the following observation.
\begin{lemma}\label{lemma:high-deg-reduce}
    Given a graph $G=(V,E)$, a density function $f(\cdot)$, and two disjoint sets $S$ and $C$ where $S$ is a low-diameter  $f(\cdot)$-dense subgraph, if there is a vertex $u\in C$ such that 
    \begin{enumerate}
        \item $|N_G(u)|= |V|-1$ or,
        \item $|N_G(u)|=|V|-2$, $|E(S\cup \{u\})|\geq f(|S|+1)$ and $diam(G[S\cup \{u\}])\leq 2$,
    \end{enumerate}
    then there is a maximum low-diameter  $f(\cdot)$-dense subgraph $S^*$ that $S\subseteq S^*\subseteq S\cup C$ and $u\in S^*$.
\end{lemma}

If $f(\cdot)$ is a hereditary-induced function, then we have more reduction rules in the following.
\begin{lemma}
\label{lemma_hereditary_reduction}
Given a graph $G=(V,E)$, a hereditary-induced function $f(\cdot)$, and two disjoint sets $S$ and $C$,
\begin{enumerate}
    \item if $S$ is not an $f(\cdot)$-dense subgraph, then there is no low-diameter $f(\cdot)$-dense subgraph $S^*$ such that $S\subseteq S^*$. 
    \item if there is a $v\in C$ such that $|E(S\cup \{v\})| < f(|S|+1)$, then there is no low-diameter $f(\cdot)$-dense subgraph $S^*$ such that $v\in S^*$.
\end{enumerate}
\end{lemma}


Notice that these reductions only hold for hereditary-induced functions. 
More reduction methods with respect to specific $f(\cdot)$ can be found based on the problem structure.
For example, in ~\citet{luo2024faster}, a number of reduction rules for the maximum $k$-defective clique problem are investigated; in ~\citet{chang2019efficient}, reduction rules for the maximum clique problem are discussed.

\subsection{Estimate Upper Bound Based On Vertex Sorting}\label{subsec:FastUB}

In this section, we introduce techniques to estimate an upper bound. 
  
\subsubsection{A Simple Bound}
\label{subsubsec:bound intuition}
To present our intuition, we reinterpret the M$f$DS problems like this --
Given a graph $G=(V,E)$ and $S\subseteq V$, find the largest vertex set $S\subseteq S^*\subseteq V$ such that $|\overline{E(S)}|+|\overline{E(S^*\setminus S)}|\le g_f(|S|)$?  (Recall that $g_f(|S|)=\binom{|S|}{2}-f(|S|)$.)
From this perspective, we can bound the size of optimal $S^*$ by under-estimating $|\overline{E(S^*\setminus S)}|$.

For example, for any vertex  $v\in C$,  define $w_S(v)=|S \setminus N(u)|$, where $w_S(v)$ represents the number of vertices in $S$ that are not neighbors to $v$. 
Then, the following algorithm gives an upper bound to the $S^*$.
\begin{enumerate}
    \item Compute $w_S(v)=|S \setminus N(v)|$ for any vertex in $C$.
    \item Order $v \in C$ by $w_S(v)$ in non-decreasing order. Denote the order as $\langle v_1,\dots,v_{|C|} \rangle$. 
    \item For any integer $k<|C|$, let us denote $\{v_1,\dots,v_k\}$ as $S_k$, and $\sum_{i=1,\dots,k}w_S(v_k)$ as $w_S(S_k)$. Find the largest index $k\in\{1,\dots,|C|\}$ such that $w_S(S_k) + \overline{E(S)}\le g_f(|S|+k)$. Then $|S|+k$ is returned as the bound.
\end{enumerate}
Note that the third step can be done by just traversing $k$ from 1 up to the first value that violates the inequality. 
\begin{lemma}
\label{lemmma_simple_bound}
    The above algorithm returns an upper bound for the maximum low-diameter $f(\cdot)$-dense graph $S^*$ that $S\subseteq S^* \subseteq S\cup C$.
\end{lemma}


\subsubsection{An Improved Sorting-based Bounding Algorithm}
\label{subsubsec:sortbound}
In fact, the correctness of the above bound relies on the fact that $w_S(S_k)+\overline{E(S)}$ underestimates the size $\overline{E(S\cup\{v_1,\dots,v_k\})}$ for any $k$. 
However, the estimation has two weak points. 
First, the edges between $S$ and set $\{v_1,\ldots, v_k\}$ are overlooked in this estimation. A second weakness of  this bound is that the two-hop constraint is also considered.
Motivated by this observation, we propose the refined algorithm, SortBound, in Alg. \ref{alg:Sort-UB} to estimate a tighter bound.

\begin{algorithm}[htb]
	\footnotesize
	\caption{A sorting-based algorithm for estimating upper bound.}\label{alg:Sort-UB}
	\KwIn{$G=(V,E)$, sets $S\subseteq V$ and $C\subseteq V$, and an oracle to $f(\cdot)$.}
	\KwOut{An upper bound size $ub$}
        SortBound($G=(V,E),f(\cdot), S,C$)\\
        \Begin{
        Greedily partition $C$ into multiple disjoint independent subsets, say $\mathcal{C}=\{\Pi_1, \Pi_2,\dots,\Pi_{\chi}\}$ where $\chi$ is the number of sets. \\
        $C'\gets \emptyset$\\
        \For{$\Pi_i$ from $1$ to $\chi$}{            
            For any $u \in \Pi_i$,  define $w_S(u)$ as $|S\setminus N(u)|$. Then, order the vertices in $\Pi_i$ by $w_S(\cdot)$ in increasing order  suppose the order as $u_1,\dots,u_{|\Pi_i|}$\\
            \While{$\Pi_{i}\neq \emptyset$}{
                    Build an empty set $R_{\sigma}$\\
                    \For{each $v$ in $\Pi_{i}$ by the sorted order}{
                        \If{$R_{\sigma}=\emptyset$, or $\forall u\in R_{\sigma}, dist_{G[S\cup C]}(u,v) > 2$}{
                            $R_{\sigma}\gets R_{\sigma}\cup \{v\}$\\
                        }
                    }
                    $\mathcal{R}\gets \mathcal{R}\cup \{R_\sigma\},\Pi_{i}\gets \Pi_{i} \setminus{R_{\sigma}}$\\
                    $\sigma\gets \sigma   +1 $                    
            }
            \For{$j$ in $\{1,2,\dots,\sigma\}$}{
                $u_j\gets \mathop{\arg\min}\limits_{u\in R_j}\ w_S(u)$\\
                $C'\gets C'\cup \{u_j\}$\\
                $w'_S(u_j) \gets w_S(u_j)+j-1$
            }
            
        }
        Sort all vertices in $C'$ by $w'_S(\cdot)$ in non-decreasing order, let $k\gets |C'|$\\
        \While{$|\overline{E}(S)|+w'_S(S_k) > g_f(|S|+k) $}{
            $k\gets k-1$
        }
        \Return{$|S|+k$}
        }
\end{algorithm}

In Alg. \ref{theorem:ub-non-heredity}, we first partition $C$ into $\chi$ independent sets, as described in line 3. 
Here, any heuristic partitioning algorithm suffices. 
But if we minimize the number of partitions $\chi$, we can obtain a tighter bound. 
Because finding the smallest $\chi$ is NP-hard (because it is equal to the graph coloring problem), we use a greedy partitioning algorithm from ~\citet{chen2021computing} to obtain such a vertex partition. In detail, suppose $C$ is initially ordered as $v_1, \dots, v_{|C|}$. We start with $\mathcal{C}$, an empty collection of sets, and greedily construct a maximal independent set from $v_1$ to $v_{|C|}$. This set is then added to $\mathcal{C}$, and we repeat the process until all vertices of $C$ are added to $\mathcal{C}$.
The time complexity of a straightforward implementation of this greedy partitioning algorithm is  $O(|C|^3)$ \citep{chen2021computing}. 

After partitioning $C$ into a collection of independent sets $\Pi_1,\ldots,\Pi_\chi$, we further partition each independent set $\Pi_i$ (where $i\in{1,...,\chi}$) into multiple subsets $R_\sigma$s such that in subgraph $G[S\cup C]$, the distance between any two vertices, say $u$ and $v$, of $R_{\sigma}$ is strictly larger than 2. 
This nested partition procedure is shown in lines 7-13 in Alg. \ref{theorem:ub-non-heredity}.
Because the diameter of the optimal solution is not larger than 2, there is at most one vertex in $R_{\sigma}$ in an optimal solution set $S^*$. 
In lines 14-17, we select the vertex with the minimum $w_S(\cdot)$ value from $R_{\sigma}$, and then add it to $C'$. 
In $C'$, we sort the vertices by $w'_S(\cdot)$ in non-decreasing order using line 18. Finally, we compute the sorting-based upper bound in lines 19-20.
One can refer to the proof of Theorem \ref{theorem:ub-non-heredity} for the correctness of the bounds.

\begin{theorem}\label{theorem:ub-non-heredity}
    Given a graph $G=(V,E)$, a growing set $S\subseteq V$ and a candidate set $C\subseteq V$, Alg. \ref{alg:Sort-UB} computes an upper bound of the size of the maximum low-diameter $f(\cdot)$-dense subgraph $S^*$ that $S\subseteq S^*\subseteq S\cup C$.
\end{theorem}


We now estimate the running time of Alg. \ref{alg:Sort-UB}.
As mentioned, the partition $\mathcal{C}$ can be obtained in time $O(|C|^3)$, that is the time of line 3 in the algorithm.

The running time of the loop from line 5 to 17 can be analyzed independently. Specifically, 
In line 6, for each $\Pi_i$, the $w_S(v)$ can be computed in $O(|S|)$ time (although we maintain $w_S(v)$ at every tree node), and the ordering is done in $O(|\Pi_i|\log{|\Pi_i|})$ time using a comparison-based sorting algorithm like quick sort. 
Therefore, the overall running time of line 6 is $O(\sum_{i=1}^{\chi}(|S|+|\Pi_i|\log |\Pi_i|))$, or simply $O(|S||C|+|C|\log |C|)$.
In lines 7 to 13, we partition each $\Pi_i$ again. The total running time of this part is $O(\sum_{i=1}^{\chi}|\Pi_i|^3)=O(|C|^3)$ because the partition still works in a greedy manner. As for lines 14 to 17, we pick a vertex from each partition $R_\sigma$ and compute a $w'_S(\cdot)$ value, which can be done in time $O(|C|^2)$.
Therefore, the total running time of the whole loop in lines 5--17 is bounded by $O(|C|^3+|S||C|)$ time.

The remaining part of the algorithm is simple: The running time of line 18 is \(O(|C|\log |C|)\) which is the complexity of sorting, and the final work in lines 19-20 of Alg. \ref{alg:Sort-UB}, which computes $k$, can be done in $O(|C|)$ time. 

In total, the overall time complexity of SortBound is $O(|C|^3+|S||C|)$. Practical speedup techniques, such as bit-parallelism ~\citep{san2011exact}, can significantly improve the pratical performance.


\subsubsection{Example and Discussion of the Bounds}

\begin{figure}[htp]
	\centering
	\includegraphics[width=0.55\textwidth]{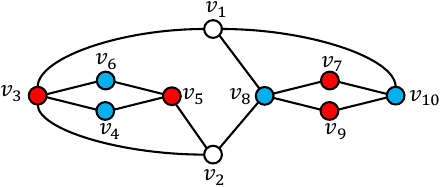}
	\caption{An example of computing sorting bound. }
	\label{fig:bound}
\end{figure}

Here we use an example to illustrate the process of computing the upper bound.
The example is illustrated for two specific $f(\cdot)$-dense graph models, $f(i)= 0.8 \binom{i}{2}$ and $f(i)=\binom{i}{2}-1$. The input graph is shown in Fig. \ref{fig:bound}. 

\begin{figure}[htp]
	\centering
	\includegraphics[width=\textwidth]{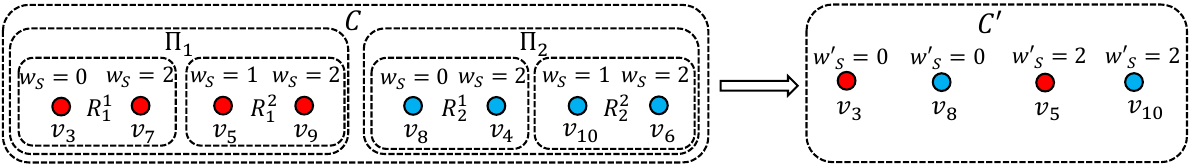}
	\caption{An example of computing sorting bound by Alg \ref{alg:Sort-UB} }
	\label{fig:bound2}
\end{figure}

In the current state of this example, we assume that $S$ includes two vertices $v_1$ and $v_2$ and $C=\{v_3,\dots,v_{10}\}$.
By line 3 in Alg. \ref{alg:Sort-UB}, suppose $C$ is partitioned into two independent sets $\Pi_1=\{v_3,v_5,v_7,v_9\}$ and $\Pi_2=\{v_4,v_6,v_8,v_{10}\}$. 
Clearly, $w_S(v_3)=w_S(v_8)=0$, $w_S(v_5)=w_S(v_{10})=1$ and $w_S(v_4)=w_S(v_6)=w_S(v_7)=w_S(v_9)=2$. 
We order $\Pi_1$ and $\Pi_2$ as $\langle v_3,v_5,v_7,v_9 \rangle$ and $\langle v_8,v_{10},v_4,v_6 \rangle$, respectively, by line 6. 
Then by lines 7-13 in Alg. \ref{alg:Sort-UB}, $\Pi_1$ and $\Pi_2$ are further partitioned into subsets $R^1_1,R^2_1$ and $R^1_2,R^2_2$, respectively. 
The partition is shown in Fig. \ref{fig:bound2}.

In this nested vertex partition, at most one vertex in a vertex set $R_j^i$ (where $i\in \{1,2\}$ and $j\in \{1,2\}$) can be moved to $S$ such that $S$ maintains to be $f(\cdot)$-dense graph.
We build up a temporary set $C'$ to keep the vertices of the smallest $w_S(\cdot)$ value in each set.
For example, in subset $R_1^1$, we move $v_3$, the vertex of smallest $w_S(\cdot)$ value in $R_1^1$, to $C'$. 
Similarly, $v_5$ in $R_1^2$, $v_8$ in $R_2^1$ and $v_{10}$ in $R_2^2$ are chosen and moved into $C'$.
Then we compute $w'_S(\cdot)$ value and obtain $w'_S(v_3)=0$,$w'_S(v_5)=2$,$w'_S(v_8)=0$ , $w'_S(v_{10})=2$. 
After a re-ordering of $C'$ by $w'_S(\cdot)$ in non-decreasing order, we obtain sequence $\langle v_3, v_8, v_5, v_{10}\rangle$.

\begin{itemize}
    \item Suppose $f(i)=0.8\binom{i}{2}$. Then we have $w'_S(S_2) + |\overline{E(S)}| \le g_f(|S|+2)$ but $w'_S(S_3) + |\overline{E(S)}| > g_f(|S|+3)$. The upper bound is $|S|+2=4$.
    \item Suppose $f(i)=\binom{i}{2}-4$. Then we have $w'_S(S_3) + |\overline{E(S)}| \le g_f(|S|+3)$ and $w'_S(S_4) + |\overline{E(S)}| > g_f(|S|+4)$. The upper bound is $|S|+3=5$. 
\end{itemize}

In fact, if $\chi=|C|$ and the distance between any two vertices of $C$ in $G[S\cup C]$ is larger than 2, the SortBound algorithm degenerates into the simple bounding algorithm described in Section \ref{subsec:FastUB}. 
On the other hand, it is known that similar sorting-based ideas are used for bounding the size of the maximum $k$-defective clique in ~\citet{chen2021computing,chang2023efficient,luo2024faster}. 
However, problem properties like two-hop distance were not exploited.

\section{Experiments}\label{sec:experiment}
All our algorithms are written in C++11 and compiled by g++ version 9.3.0 with the -Ofast flag. 
All experiments are conducted on a machine with an Intel(R) Xeon(R) Platinum 8360Y CPU @ 2.40GHz and an Ubuntu 22.04 operating system. Hyper-threading and turbo are disabled for steady clock frequency.  

\subsection{Benchmark Algorithms}
In this section, we show the experimental results of the 6 algorithms.
\begin{enumerate}
    \item \textbf{Deg-BnB}: The decomposition algorithm described in Alg. \ref{Framework}, which uses the heuristic algorithm from Alg. \ref{alg:Heualg} as the initial heuristic. It employs the degeneracy ordering from Section \ref{subsubsec:degeneracyordering} for graph decomposition and the branch-and-bound algorithm with the sorting-based approach from Alg. \ref{alg-branch-bound} as the ExactSearch method.
    \item \textbf{TwoDeg-BnB}: A variant of Deg-BnB that uses the \textit{two-hop degeneracy ordering} from Section \ref{subsubsubsec:weak-degen} for graph decomposition. Other components remain the same as Deg-BnB.
    \item \textbf{BnB}: The branch-and-bound algorithm from Alg. \ref{alg-branch-bound} without graph decomposition. It uses the heuristic algorithm from Alg. \ref{alg:Heualg} to obtain an initial lower bound ($lb$), and sets \( S = \emptyset \) and \( C = V \).
    \item \textbf{MIP}: The Gurobi solver using the MIP-$f$D formulation presented in Section \ref{section_mip}.
    \item \textbf{Deg-MIP}: A variant of Deg-BnB that replaces the ExactSearch algorithm with the Gurobi solver using the MIP-$f$D formulation. All other elements remain the same as Deg-BnB.
    \item \textbf{TwoDeg-MIP}: A variant of TwoDeg-BnB that uses the Gurobi solver with the MIP-$f$D formulation as ExactSearch, while keeping all other components the same as TwoDeg-BnB.
\end{enumerate}

The source codes of these algorithms are publicly available at \url{https://github.com/cy-Luo000/MDSL}.

\subsection{Datasets}
We ran the algorithms on large graph collections derived from real-world applications. 
The dataset consists of 139 undirected graphs, with up to \(5.87 \times 10^7\) vertices and \(1.06 \times 10^8\) undirected edges, sourced from the Network Data Repository\footnote{\url{https://networkrepository.com/}}. The graphs represent various domains, including social networks, biological networks, collaboration networks, and more.


We test two specific $f(\cdot)$-dense functions.
\begin{enumerate}
    \item The first $f(\cdot)$-dense function is defined as $f(i)=\gamma\binom{i}{2}$ where $\gamma$ are chosen from $\{0.99, 0.95, 0.90,0.85\}$. 
    \item The second $f(\cdot)$-dense function is defined as $f(i)=\binom{i}{2}-s$ where $s$ is chosen from $\{1,3\}$. This model represents the so-called $s$-defective clique.
\end{enumerate}
Note that, $f(i)=\gamma\binom{i}{2}$ is not a hereditary-induced function, whereas $f(i)=\binom{i}{2}-s$ is a hereditary-induced function. It is clear that the $f(\cdot)$-dense where $f(i)=\gamma\binom{i}{2}$ represents the $\gamma$-quasi clique, while $f(i)=\binom{i}{2}-s$ represents the $s$-defective clique.

\subsection{The Empirical Running Time}\label{subsec:general-experiment}

\begin{figure}[htb]
	\centering  
    \scalebox{0.9}{  
    \begin{tabular}{cc} 
	\subfigure[$\gamma=0.99$]{
		\begin{minipage}[b]{0.45\textwidth}
			\includegraphics[width=1\textwidth]{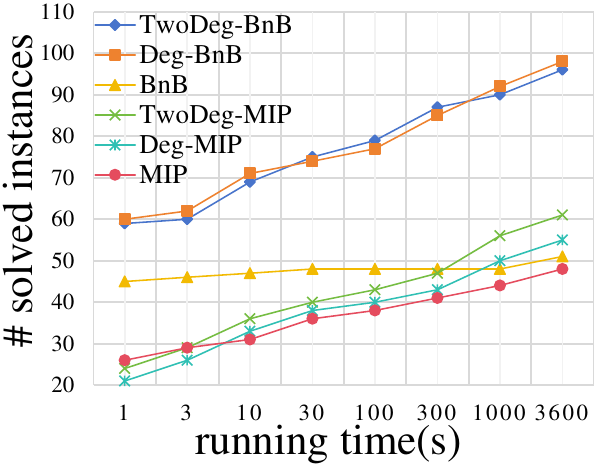} 
		\end{minipage}
		\label{subfig:overall-0.99}
	}
        \subfigure[$\gamma=0.95$]{
            \begin{minipage}[b]{0.45\textwidth}
            \includegraphics[width=1\textwidth]{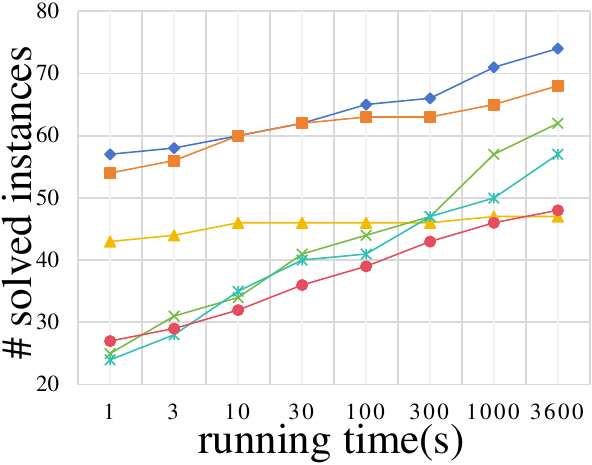}
            \end{minipage}
        \label{subfig:overall-0.95}
        }
        \\
         \subfigure[$\gamma=0.90$]{
            \begin{minipage}[b]{0.45\textwidth}
            \includegraphics[width=1\textwidth]{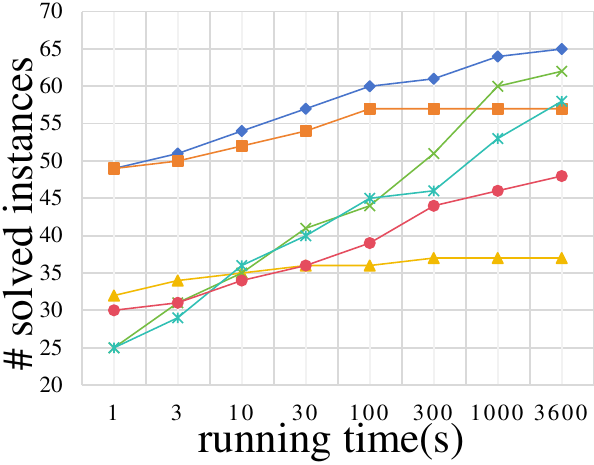}
            \end{minipage}
        \label{subfig:overall-0.90}
        }	 
	\subfigure[$\gamma=0.85$]{
		\begin{minipage}[b]{0.45\textwidth}
			\includegraphics[width=1\textwidth]{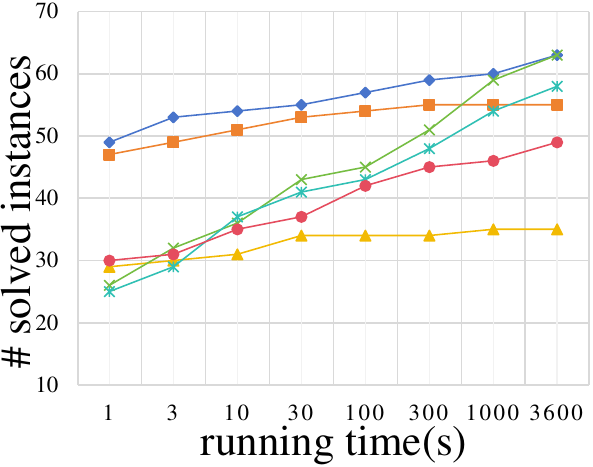} 
		\end{minipage}
		\label{subfig:overall-0.85}
	}
    \end{tabular}
    }   
    \caption{Overall performance evaluation of M$f$DS algorithms for $f(i)=\gamma\binom{i}{2}$ with $\gamma=\{0.99,\ 0.95,\ 0.90,\ 0.85\}$.}
	\label{fig:overall-pic-qc}
\end{figure}

\begin{figure}[htb]
	\centering
        \scalebox{0.85}{
            \begin{tabular}{cc} 
        	\subfigure[$s=1$]{
        		\begin{minipage}[b]{0.45\textwidth}
        			\includegraphics[width=1\textwidth]{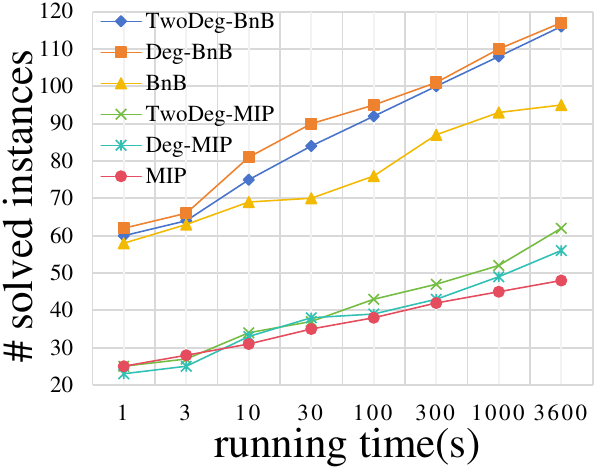} 
        		\end{minipage}
        		\label{subfig:overall-1}
        	}
                 \subfigure[$s=3$]{
                    \begin{minipage}[b]{0.45\textwidth}
                    \includegraphics[width=1\textwidth]{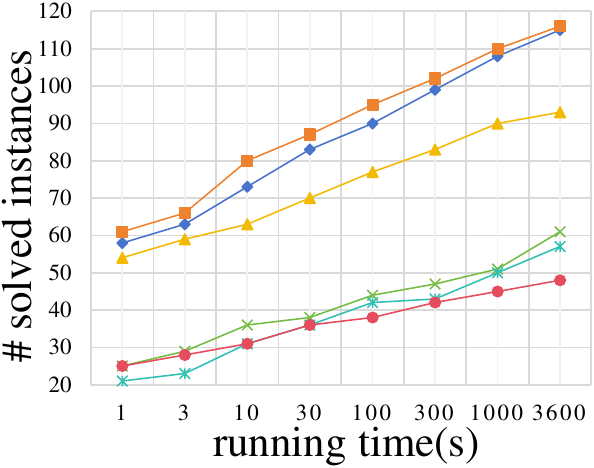}
                    \end{minipage}
                \label{subfig:overall-3}
                }
            \end{tabular}
        }
	\caption{Overall performance evaluation of M$f$DS algorithms for $f_2(i)=\binom{i}{2}-s$ with $s=\{1,3\}$.}
	\label{fig:overall-pic-kdc}
\end{figure}

In Fig. \ref{fig:overall-pic-qc} and \ref{fig:overall-pic-kdc}, we present the number of instances solved by each algorithm across different time frames. The upper time limit for each run is set to 3600 seconds (1 hour). 
In general, the combinatorial algorithms TwoDeg-BnB and Deg-BnB seem to be the most competitive.
Given \( f(i) = \gamma \binom{i}{2} \) where $\gamma$ is $0.99$ or $0,95$,  and \( f(i) = \binom{i}{2} - s \) where $s=1$ or $3$, TwoDeg-BnB and Deg-BnB fully outperform the other four algorithms across various time frames. 
Given \( f(i) = \gamma \binom{i}{2} \) where $\gamma$ is $0.90$ or $0,85$, after running more than 1000 seconds, TwoDeg-MIP and Deg-MIP solve a similar number of instances as TwoDeg-BnB and Deg-BnB.
However, both branch-and-bound and MIP-based algorithms benefit from the integration into the decomposition framework, resulting in improved computation times. 
Within one hour, the algorithm with decomposition solved nearly twice as many instances optimally as the algorithm without decomposition.
However, it remains difficult to determine which graph decomposition ordering is superior. 


We further present detailed results for graphs containing at least 10,000 vertices, where at least one algorithm is able to solve the graph to optimality for each value of \( \gamma \) and \( s \). In total, 15 graphs meet these criteria. We report the detailed results of these 15 graphs in Table \ref{tab:detailed-cmp-qc} for \( f(i) = \gamma \binom{i}{2} \) and Table \ref{tab:detailed-cmp-kdc} for \( f(i) = \binom{i}{2}-s \). 
The optimal solution size is shown in the column \textit{opt}, and the remaining columns display the empirical running times of each algorithm. 
The label \textit{-} indicates that the algorithm exceeded the time limit. 


In Table \ref{tab:detailed-cmp-qc}, the algorithms with graph decomposition achieve a speedup of up to 2-3 orders of magnitude. There are 39 instances where TwoDeg-BnB or Deg-BnB solve to optimality, but BnB fails. 
In contrast, all instances solvable by BnB can also be solved by either TwoDeg-MIP or Deg-MIP. 
Combinatorial algorithms including TwoDeg-BnB, Deg-BnB and BnB mostly outperform their MIP-based counterparts (TwoDeg-MIP, Deg-MIP, and MIP). 
For the MIP solver, the inclusion of graph decomposition significantly accelerates the algorithm. In fact, while MIP fails to solve any of these graphs on its own, with graph decomposition, it sometimes outperforms pure combinatorial algorithms, as seen with the web-sk-2005 graph for \( \gamma = 0.95, 0.9, \) and \( 0.85 \).

For the case of \( f(i) = \binom{i}{2} - s \) where results are given in Table \ref{tab:detailed-cmp-kdc}, we observe similar results. However, for this \( f(\cdot) \)-dense function, the pure combinatorial BnB algorithm seems to be considerably more efficient than MIP.

 {
   
    \renewcommand{\arraystretch}{0.8}
    \scriptsize
    \begin{longtable}{ccccccccc}
    \caption{
    The running times (in seconds) of TwoDeg-BnB, Deg-BnB, BnB, TwoDeg-MIP, Deg-MIP, MIP on 15 large real-world graphs for $f(i)=\gamma\binom{i}{2}$ and $\gamma=0.99,\ 0.95,\ 0.90,\ 0.85$.
    }        
    \label{tab:detailed-cmp-qc}\\
    \toprule
    \textbf{\makecell{Graph\\ information}} & $\gamma$ & \textbf{opt} & \textbf{\makecell{TwoDeg-\\ BnB}} & \textbf{\makecell{Deg-\\BnB}} & \textbf{BnB} & \textbf{\makecell{TwoDeg-\\ MIP}} & \textbf{\makecell{Deg-\\MIP}} & \textbf{MIP} \\
    \midrule
    scc\_infect-dublin & 0.99  & 86    & -   & \textbf{0.54 } & -   & -   & -   & - \\
    |V|=10972 & 0.95  & 94    & -   & -   & -   & \textbf{2093.18 } & -   & - \\
    |E|=175573 & 0.9   & 101   & -   & -   & -   & \textbf{2035.23 } & 2897.83  & - \\
    $d_G=$83, $td_G=$219 & 0.85  & 108   & -   & -   & -   & \textbf{1967.59 } & 3436.76  & - \\
    \midrule
    web-indochina-2004 & 0.99  & 50    & 0.25  & \textbf{0.20 } & 12.10  & 168.85  & 346.24  & - \\
    |V|=11358 & 0.95  & 51    & \textbf{0.47 } & 0.89  & -   & 161.70  & 199.26  & - \\
    |E|=47606 & 0.9   & 51    & \textbf{30.99 } & -   & -   & 178.98  & 345.32  & - \\
    $d_G=$49, $td_G=$199 & 0.85  & 51    & -   & -   & -   & \textbf{207.01 } & 313.21  & - \\
    \midrule
    web-BerkStan & 0.99  & 29    & \textbf{0.03 } & \textbf{0.03 } & -   & 61.21  & 6.65  & - \\
    |V|=12305 & 0.95  & 29    & 0.05  & \textbf{0.03 } & -   & 59.44  & 3.19  & - \\
    |E|=19500 & 0.9   & 30    & 0.03  & \textbf{0.01 } & -   & 40.68  & 5.28  & - \\
    $d_G=$28, $td_G=$59 & 0.85  & 31    & 0.03  & \textbf{0.02 } & -   & 55.14  & 5.03  & - \\
    \midrule
    web-webbase-2001 & 0.99  & 33    & \textbf{0.33 } & 4.06  & -   & 700.17  & -   & - \\
    |V|=16062 & 0.95  & 33    & \textbf{0.27 } & 5.79  & -   & 709.22  & -   & - \\
    |E|=25593 & 0.9   & 34    & \textbf{0.31 } & -   & -   & 694.47  & -   & - \\
    $d_G=$32, $td_G=$1679 & 0.85  & 35    & \textbf{0.21 } & -   & -   & 651.36  & -   & - \\
    \midrule
    scc\_retweet-crawl & 0.99  & 21    & 0.12  & \textbf{0.10 } & -   & 1020.72  & 1608.18  & - \\
    |V|=17151 & 0.95  & 23    & \textbf{0.17 } & -   & -   & 597.08  & 1272.85  & - \\
    |E|=24015 & 0.9   & 26    & \textbf{1.32 } & -   & -   & 550.35  & 620.96  & - \\
    $d_G=$19, $td_G=$195 & 0.85  & 28    & \textbf{124.08 } & -   & -   & 365.67  & 659.36  & - \\
    \midrule
    ca-CondMat & 0.99  & 26    & -   & \textbf{0.93 } & -   & -   & -   & - \\
    |V|=21363 & 0.95  & 28    & -   & -   & -   & \textbf{2679.19 } & -   & - \\
    |E|=91286 & 0.9   & 30    & \textbf{22.54 } & -   & -   & 2489.77  & -   & - \\
    $d_G=$25, $td_G=$279 & 0.85  & 31    & -   & -   & -   & \textbf{3307.37 } & -   & - \\
    \midrule
    tech-p2p-gnutella & 0.99  & 4     & 0.75  & \textbf{0.55 } & -   & 1663.14  & 1405.14  & - \\
    |V|=62561 & 0.95  & 4     & 0.71  & \textbf{0.53 } & -   & 2280.71  & 1139.19  & - \\
    |E|=147878 & 0.9   & 5     & 0.79  & \textbf{0.58 } & -   & 834.06  & 706.13  & - \\
    $d_G=$6, $td_G=$95 & 0.85  & 5     & 0.77  & \textbf{0.59 } & -   & 1310.10  & 1627.57  & - \\
    \midrule
    rec-amazon & 0.99  & 5     & 0.20  & \textbf{0.18 } & -   & 392.30  & 176.02  & - \\
    |V|=91813 & 0.95  & 5     & 0.19  & \textbf{0.18 } & -   & 414.68  & 143.18  & - \\
    |E|=125704 & 0.9   & 6     & 0.19  & \textbf{0.16 } & -   & 151.62  & 89.83  & - \\
    $d_G=$4, $td_G=$8 & 0.85  & 6     & 0.18  & \textbf{0.16 } & -   & 152.20  & 89.32  & - \\
    \midrule
    web-sk-2005 & 0.99  & 84    & \textbf{1.87 } & -   & -   & 915.49  & 1137.96  & - \\
    |V|=121422 & 0.95  & 84    & 1737.03  & -   & -   & 1350.16  & \textbf{1219.74 } & - \\
    |E|=334419 & 0.9   & 86    & -   & -   & -   & \textbf{818.19 } & 1320.09  & - \\
    $d_G=$81, $td_G=$590 & 0.85  & 89    & -   & -   & -   & \textbf{774.55 } & 957.47  & - \\
    \midrule
    web-uk-2005 & 0.99  & 501   & \textbf{634.49 } & -   & -   & 768.52  & -   & - \\
    |V|=129632 & 0.95  & 501   & \textbf{424.08 } & -   & -   & 836.24  & -   & - \\
    |E|=11744049 & 0.9   & 501   & \textbf{634.16 } & -   & -   & 666.69  & -   & - \\
    $d_G=$499, $td_G=$850 & 0.85  & 501   & 709.77  & -   & -   & \textbf{631.67 } & -   & - \\
    \midrule
    web-arabic-2005 & 0.99  & 102   & \textbf{8.83 } & -   & -   & 830.86  & -   & - \\
    |V|=163598 & 0.95  & 104   & -   & -   & -   & \textbf{726.84 } & 3267.47  & - \\
    |E|=1747269 & 0.9   & 107   & -   & -   & -   & \textbf{437.76 } & -   & - \\
    $d_G=$101, $td_G=$1102 & 0.85  & 111   & -   & -   & -   & \textbf{554.77 } & -   & - \\
    \midrule
    ca-MathSciNet & 0.99  & 25    & 11.81  & \textbf{6.91 } & -   & -   & -   & - \\
    |V|=332689 & 0.95  & 25    & 10.62  & \textbf{5.94 } & -   & -   & -   & - \\
    |E|=820644 & 0.9   & 26    & \textbf{10.37 } & 11.36  & -   & -   & -   & - \\
    $d_G=$24, $td_G=$496 & 0.85  & 27    & \textbf{170.16 } & -   & -   & -   & -   & - \\
    \midrule
    inf-roadNet-PA & 0.99  & 4     & 3.14  & \textbf{2.25 } & -   & -   & -   & - \\
    |V|=1087562 & 0.95  & 4     & 2.73  & \textbf{2.31 } & -   & -   & -   & - \\
    |E|=1541514 & 0.9   & 4     & 2.94  & \textbf{2.15 } & -   & -   & -   & - \\
    $d_G=$3, $td_G=$9 & 0.85  & 4     & 2.86  & \textbf{2.28 } & -   & -   & -   & - \\
    \midrule
    inf-roadNet-CA & 0.99  & 4     & 5.64  & \textbf{3.96 } & -   & -   & -   & - \\
    |V|=1957027 & 0.95  & 4     & 5.02  & \textbf{4.46 } & -   & -   & -   & - \\
    |E|=2760388 & 0.9   & 4     & 5.28  & \textbf{3.91 } & -   & -   & -   & - \\
    $d_G=$3, $td_G=$12 & 0.85  & 4     & 5.15  & \textbf{4.57 } & -   & -   & -   & - \\
    \midrule
    inf-road-usa & 0.99  & 4     & 70.78  & \textbf{40.80 } & -   & -   & -   & - \\
    |V|=23947347 & 0.95  & 4     & 74.46  & \textbf{41.31 } & -   & -   & -   & - \\
    |E|=28854312 & 0.9   & 4     & 79.18  & \textbf{46.12 } & -   & -   & -   & - \\
    $d_G=$3, $td_G=$9 & 0.85  & 4     & 82.86  & \textbf{46.50 } & -   & -   & -   & - \\
    \bottomrule
    \end{longtable}
    }

   {
    \renewcommand{\arraystretch}{0.8}
    \scriptsize
   
\begin{longtable}{ccccccccc}
     \caption{
The running times (in seconds) of TwoDeg-BnB, Deg-BnB, BnB, TwoDeg-MIP, Deg-MIP, MIP on 15 large real-world graphs when $f(i)=\binom{i}{2}-s$ and $s=1,\ 3$.  
        }
    \label{tab:detailed-cmp-kdc} \\
     \toprule
   \textbf{\makecell{Graph\\ information}} & $s$ & \textbf{opt} & \textbf{\makecell{TwoDeg-\\ BnB}} & \textbf{\makecell{Deg-\\BnB}} & \textbf{BnB} & \textbf{\makecell{TwoDeg-\\ MIP}} & \textbf{\makecell{Deg-\\MIP}} & \textbf{MIP} \\
    \midrule
    scc\_infect-dublin & 1     & 84    & 0.83  & \textbf{0.43 } & 1.62  & 3314.12  & -   & - \\
    $d_G=$83, $td_G=$219 & 3     & 84    & 1.14  & \textbf{0.32 } & 2.10  & -   & -   & - \\
    \midrule
    web-indochina-2004 & 1     & 50    & \textbf{0.18 }  & \textbf{0.18 } & 5.04  & 221.47  & 358.84  & - \\
    $d_G=$49, $td_G=$199 & 3     & 50    & \textbf{0.17 }  & 0.18  & 5.41  & 174.66  & 430.02  & - \\
    \midrule
    web-BerkStan & 1     & 29    & \textbf{0.03 } & \textbf{0.03 } & 0.86  & 64.58  & 6.50  & - \\
    $d_G=$28, $td_G=$59 & 3     & 29    & 0.05  & \textbf{0.02 } & 1.13  & 45.59  & 6.76  & - \\
    \midrule
    web-webbase-2001 & 1     & 33    & \textbf{0.33 } & 4.56  & 2.19  & 1516.71  & -   & - \\
    $d_G=$32, $td_G=$1679 & 3     & 33    & \textbf{0.25 } & 7.03  & 13.04  & 1538.96  & -   & - \\
    \midrule
    scc\_retweet-crawl & 1     & 21    & \textbf{0.09 }  & \textbf{0.09 } & 0.10  & 1289.17  & 1654.74  & - \\
    $d_G=$19, $td_G=$195 & 3     & 21    & 0.16  & \textbf{0.09 } & 0.13  & 1107.25  & 2078.58  & - \\
    \midrule
    ca-CondMat & 1     & 26    & 1.00  & \textbf{0.90 } & 11.46  & 3321.07  & -   & - \\
    $d_G=$25, $td_G=$279 & 3     & 26    & 1.00  & \textbf{0.85 } & 17.38  & -   & -   & - \\
    \midrule
    tech-p2p-gnutella & 1     & 5     & 0.70  & \textbf{0.55 } & 272.72  & 1496.14  & 1292.12  & - \\
    $d_G=$6, $td_G=$95 & 3     & 5     & 0.72  & \textbf{0.63 } & 339.75  & 1772.54  & 1931.10  & - \\
    \midrule
    rec-amazon & 1     & 6     & \textbf{0.08 }  & 0.09  & 264.28  & 59.22  & 165.39  & - \\
    $d_G=$4, $td_G=$8 & 3     & 6     & 0.08  & \textbf{0.06 } & -   & 8.60  & 87.80  & - \\
    \midrule
    web-sk-2005 & 1     & 82    & \textbf{0.80 } & 1.36  & -   & 1195.99  & 2419.91  & - \\
    $d_G=$81, $td_G=$590 & 3     & 83    & \textbf{0.84 }  & 1.73  & -   & 1056.36  & 2317.35  & - \\
    \midrule
    web-uk-2005 & 1     & 500   & 391.19  & \textbf{30.72 } & -   & 1873.15  & -   & - \\
    $d_G=$499, $td_G=$850 & 3     & 500   & 365.31  & \textbf{43.03 } & -   & 1880.04  & -   & - \\
    \midrule
    web-arabic-2005 & 1     & 102   & \textbf{2.79 } & 4.91  & -   & 1671.32  & -   & - \\
    $d_G=$101, $td_G=$1102 & 3     & 102   & \textbf{2.91 } & 4.90  & -   & 1502.21  & -   & - \\
    \midrule
    ca-MathSciNet & 1     & 25    & 7.92  & \textbf{6.59 } & -   & -   & -   & - \\
    $d_G=$24, $td_G=$496 & 3     & 25    & 7.62  & \textbf{7.07 } & -   & -   & -   & - \\
    \midrule
    inf-roadNet-PA & 1     & 4     & 1.82  & \textbf{1.57 } & -   & -   & -   & - \\
    $d_G=$3, $td_G=$9 & 3     & 5     & 1.54  & \textbf{1.29 } & -   & 3136.69  & 3374.48  & - \\
    \midrule
    inf-roadNet-CA & 1     & 4     & 3.40  & \textbf{2.79 } & -   & -   & -   & - \\
    $d_G=$3, $td_G=$12 & 3     & 5     & 2.90  & \textbf{2.24 } & -   & -   & -   & - \\
    \midrule
    inf-road-usa & 1     & 4     & 44,55  & \textbf{37.75 } & -   & -   & -   & - \\
    $d_G=$3, $td_G=$9 & 3     & 5     & 43.82  & \textbf{32.63 } & -   & -   & -   & - \\
    \bottomrule
    \end{longtable}
}

\subsection{Effectiveness of Upper Bound}
 
In this section, we conduct experiments to evaluate the effectiveness of the proposed upper bounding algorithms using the 15 graphs mentioned above. In Table \ref{tab:detailed-ub-mqc} and \ref{tab:detailed-ub-kdc} , we present the results for \( f(i) = \gamma \binom{i}{2} \) and \(f(i)=\binom{i}{2}-s\), respectively. 
In columns \textit{LP}, \textit{Simple}, and \textit{Sortbound}, we report the upper bounds of the M$f$DS for the input graph, computed using the linear relaxation of MIP-$f$D (see Section \ref{section_mip}), the simple sorting bound algorithm (see Section \ref{subsubsec:bound intuition}), and the SortBound bounding algorithm (see Section \ref{subsubsec:sortbound}), respectively. 
We also report the running time in the column \textit{Time} and the number of tree nodes in the column \textit{Nodes} for the algorithms TwoDeg-BnB and TwoDeg-BnB/ub. The algorithm TwoDeg-BnB/ub is a modified version of TwoDeg-BnB that excludes the SortBound algorithm described in Section \ref{subsubsec:sortbound}.

In terms of the tightness of the bounds, we observe that, in most cases, the LP relaxation provides the tightest upper bound, followed by our SortBound algorithm, with the Simple bound algorithm being the least tight. In a few instances, SortBound yields a better upper bound than the LP relaxation. 
However, while the LP relaxation provides the tightest bounds, it fails to scale to larger instances, such as \textit{web-uk-2005} and other large graphs, where the linear programming solver cannot compute the upper bounds. 
In contrast, both SortBound and the Simple bound algorithms can handle all large instances efficiently. 
Therefore, we conclude that SortBound obtains the best balance between running time and tightness. It provides bounds tight enough to significantly prune the search tree, thereby speeding up the search algorithm. For example, for the graph \textit{inf-roadNet-CA} with \( f(\cdot) = 0.95 \binom{i}{2} \), SortBound reduces the search tree size from almost \( 1.4 \times 10^7 \) to less than \( 1 \times 10^3 \).

As for the function \( f(i) = \binom{i}{2} - s \) with \( s = 1 \) and \( 3 \), SortBound is still more effective than LP. 
In some extreme cases, such as \textit{ca-MathSciNet} with \( f(\cdot) = \binom{i}{2} - 3 \), the speedup achieved by using SortBound is as large as 100x compared to the algorithm without this bound.


{
    \setlength{\tabcolsep}{0.6mm}
    \captionsetup[longtable]{font=normalsize}
    \renewcommand{\arraystretch}{0.8}
    \scriptsize

   
    \begin{longtable}{ccc|ccc|cc|cc}

        \caption{
    The comparison of different bounds on 15 large real-world graphs for $f(i)=\gamma\binom{i}{2}$ where $\gamma=0.99,\ 0.95,\ 0.90,\ 0.85$.  
        }
    \label{tab:detailed-ub-mqc} \\
    \toprule
    \multirow{2}[4]{*}{\textbf{Graph}} & \multirow{2}[4]{*}{\textbf{$\gamma$}} & \multirow{2}[4]{*}{\textbf{opt}} & \multirow{2}[4]{*}{\textbf{LP}} & \multirow{2}[4]{*}{\textbf{Simple}} & \multirow{2}[4]{*}{\textbf{SortBound}} & \multicolumn{2}{c|}{\textbf{Time(s)}} & \multicolumn{2}{c}{\textbf{Nodes($10^3$)}} \\
\cline{7-8} \cline{9-10}         &       &       &       &       &       & \textbf{\makecell{TwoDeg-\\BnB}} & \textbf{\makecell{TwoDeg-\\BnB/ub}} & \textbf{\makecell{TwoDeg-\\BnB}} & \textbf{\makecell{TwoDeg-\\BnB/ub}} \\
    \hline
    scc\_infect-dublin & 0.99  & 86    & \textbf{94} & 220   & 168   & -   & -   & -   & - \\
    |V|=10972 & 0.95  & 94    & \textbf{97} & 220   & 220   & -   & -   & -   & - \\
    |E|=175573 & 0.9   & 101   & \textbf{103} & 220   & 220   & -   & -   & -   & - \\
    $d_G=$83, $td_G=$- & 0.85  & 108   & \textbf{109} & 220   & 220   & -   & -   & -   & - \\
    \hline
    web-indochina-2004 & 0.99  & 50    & \textbf{50} & 200   & 54    & \textbf{0.25 } & -   & \textbf{0.01 } & - \\
    |V|=11358 & 0.95  & 51    & \textbf{51} & 200   & 112   & \textbf{0.47 } & -   & \textbf{15.27 } & - \\
    |E|=47606 & 0.9   & 51    & \textbf{51} & 200   & 158   & \textbf{30.99 } & -   & \textbf{3920.40 } & - \\
    $d_G=$49, $td_G=$199 & 0.85  & 51    & \textbf{51} & 200   & 162   & -   & -   & -   & - \\
    \hline
    web-BerkStan & 0.99  & 29    & \textbf{29} & 60    & 33    & \textbf{0.03 } & -   & \textbf{0.06 } & - \\
    |V|=12305 & 0.95  & 29    & \textbf{29} & 60    & 33    & \textbf{0.05 } & -   & \textbf{0.24 } & - \\
    |E|=19500 & 0.9   & 30    & \textbf{30} & 60    & 33    & \textbf{0.03 } & -   & \textbf{0.68 } & - \\
    $d_G=$28, $td_G=$59 & 0.85  & 31    & \textbf{31} & 60    & 33    & \textbf{0.03 } & -   & \textbf{0.51 } & - \\
    \hline
    web-webbase-2001 & 0.99  & 33    & \textbf{33} & 1680  & 38    & \textbf{0.33 } & -   & \textbf{0.10 } & - \\
    |V|=16062 & 0.95  & 33    & \textbf{33} & 1680  & 39    & \textbf{0.27 } & -   & \textbf{0.16 } & - \\
    |E|=25593 & 0.9   & 34    & \textbf{34} & 1680  & 39    & \textbf{0.31 } & -   & \textbf{0.56 } & - \\
    $d_G=$32, $td_G=$1679 & 0.85  & 35    & \textbf{36} & 1680  & 61    & \textbf{0.21 } & -   & \textbf{1.55 } & - \\
    \hline
    scc\_retweet-crawl & 0.99  & 21    & 26    & 196   & \textbf{25} & \textbf{0.12 } & -   & \textbf{0.67 } & - \\
    |V|=17151 & 0.95  & 23    & \textbf{27} & 196   & 77    & \textbf{0.17 } & -   & \textbf{5.95 } & - \\
    |E|=24015 & 0.9   & 26    & \textbf{29} & 196   & 112   & \textbf{1.32 } & -   & \textbf{139.65 } & - \\
    $d_G=$19, $td_G=$195 & 0.85  & 28    & \textbf{30} & 196   & 145   & \textbf{124.08 } & -   & \textbf{5006.13 } & - \\
    \hline
    ca-CondMat & 0.99  & 26    & \textbf{28} & 280   & 30    & -   & -   & -   & - \\
    |V|=21363 & 0.95  & 28    & \textbf{29} & 280   & 142   & -   & -   & -   & - \\
    |E|=91286 & 0.9   & 30    & \textbf{30} & 280   & 206   & \textbf{22.54 } & -   & \textbf{1271.69 } & - \\
    $d_G=$25, $td_G=$279 & 0.85  & 31    & \textbf{31} & 280   & 270   & -   & -   & -   & - \\
    \hline
    tech-p2p-gnutella & 0.99  & 4     & 7     & 96    & \textbf{5} & \textbf{0.75 } & -   & \textbf{0.01 } & - \\
    |V|=62561 & 0.95  & 4     & 7     & 96    & \textbf{5} & \textbf{0.71 } & -   & \textbf{0.01 } & - \\
    |E|=147878 & 0.9   & 5     & 8     & 96    & \textbf{7} & \textbf{0.79 } & -   & \textbf{0.11 } & - \\
    $d_G=$6, $td_G=$95 & 0.85  & 5     & \textbf{8} & 96    & 9     & \textbf{0.77 } & -   & \textbf{0.14 } & - \\
    \hline
    rec-amazon & 0.99  & 5     & \textbf{5} & 9     & \textbf{5} & \textbf{0.20 } & 0.29  & \textbf{0.00 } & 181.90  \\
    |V|=91813 & 0.95  & 5     & \textbf{5} & 9     & \textbf{5} & \textbf{0.19 } & 0.27  & \textbf{0.00 } & 181.90  \\
    |E|=125704 & 0.9   & 6     & \textbf{6} & 9     & 7     & 0.19  & \textbf{0.17 } & \textbf{0.00 } & 62.63  \\
    $d_G=$4, $td_G=$8 & 0.85  & 6     & \textbf{6} & 9     & 8     & \textbf{0.18 } & 0.20  & \textbf{0.01 } & 58.69  \\
    \hline
    web-sk-2005 & 0.99  & 84    & \textbf{84} & 591   & 114   & \textbf{1.87 } & -   & \textbf{17.67 } & - \\
    |V|=121422 & 0.95  & 84    & \textbf{84} & 591   & 135   & \textbf{1737.03 } & -   & \textbf{120249.90 } & - \\
    |E|=334419 & 0.9   & 86    & \textbf{87} & 591   & 154   & -   & -   & -   & - \\
    $d_G=$81, $td_G=$590 & 0.85  & 89    & \textbf{90} & 591   & 180   & -   & -   & -   & - \\
    \hline
    web-uk-2005 & 0.99  & 501   & -   & 851   & \textbf{501} & \textbf{634.49 } & -   & \textbf{32.59 } & - \\
    |V|=129632 & 0.95  & 501   & -   & 851   & \textbf{501} & \textbf{424.08 } & -   & \textbf{0.00 } & - \\
    |E|=11744049 & 0.9   & 501   & -   & 851   & \textbf{501} & \textbf{634.16 } & -   & \textbf{0.00 } & - \\
    $d_G=$499, $td_G=$850 & 0.85  & 501   & -   & 851   & \textbf{501} & \textbf{709.77 } & -   & \textbf{0.00 } & - \\
    \hline
    web-arabic-2005 & 0.99  & 102   & \textbf{102} & 1103  & 146   & \textbf{8.83 } & -   & 63.81  & - \\
    |V|=163598 & 0.95  & 104   & \textbf{105} & 1103  & 258   & -   & -   & -   & - \\
    |E|=1747269 & 0.9   & 107   & \textbf{108} & 1103  & 258   & -   & -   & -   & - \\
    $d_G=$101, $td_G=$1102 & 0.85  & 111   & \textbf{112} & 1103  & 411   & -   & -   & -   & - \\
    \hline
    ca-MathSciNet & 0.99  & 25    & -   & 497   & \textbf{29} & \textbf{11.81 } & -   & \textbf{0.13 } & - \\
    |V|=332689 & 0.95  & 25    & -   & 497   & \textbf{74} & \textbf{10.62 } & -   & \textbf{1.13 } & - \\
    |E|=820644 & 0.9   & 26    & -   & 497   & \textbf{168} & \textbf{10.37 } & -   & \textbf{145.85 } & - \\
    $d_G=$24, $td_G=$496 & 0.85  & 27    & -   & 497   & \textbf{206} & \textbf{170.16 } & -   & \textbf{15415.56 } & - \\
    \hline
    inf-roadNet-PA & 0.99  & 4     & -   & 10    & \textbf{4} & \textbf{3.14 } & 5.56  & \textbf{0.00 } & 7972.09  \\
    |V|=1087562 & 0.95  & 4     & -   & 10    & \textbf{4} & \textbf{2.73 } & 5.76  & \textbf{0.00 } & 7972.09  \\
    |E|=1541514 & 0.9   & 4     & -   & 10    & \textbf{5} & \textbf{2.94 } & 5.09  & \textbf{0.00 } & 7972.09  \\
    $d_G=$3, $td_G=$9 & 0.85  & 4     & -   & 10    & \textbf{7} & \textbf{2.86 } & 5.82  & \textbf{0.00 } & 7972.09  \\
    \hline
    inf-roadNet-CA & 0.99  & 4     & -   & 13    & \textbf{5} & \textbf{5.64 } & 10.25  & \textbf{0.01 } & 14138.82  \\
    |V|=1957027 & 0.95  & 4     & -   & 13    & \textbf{5} & \textbf{5.02 } & 9.97  & \textbf{0.01 } & 14138.82  \\
    |E|=2760388 & 0.9   & 4     & -   & 13    & \textbf{7} & \textbf{5.28 } & 9.16  & \textbf{0.02 } & 14138.82  \\
    $d_G=$3, $td_G=$12 & 0.85  & 4     & -   & 13    & \textbf{7} & \textbf{5.15 } & 10.64  & \textbf{0.02 } & 14138.82  \\
    \hline
    inf-road-usa & 0.99  & 4     & -   & 10    & \textbf{4} & \textbf{70.78 } & 103.02  & \textbf{0.01 } & 69471.02  \\
    |V|=23947347 & 0.95  & 4     & -   & 10    & \textbf{4} & \textbf{74.46 } & 86.16  & \textbf{0.01 } & 69471.02  \\
    |E|=28854312 & 0.9   & 4     & -   & 10    & \textbf{5} & \textbf{79.18 } & 101.55  & \textbf{0.02 } & 69471.02  \\
    $d_G=$3, $td_G=$9 & 0.85  & 4     & -   & 10    & \textbf{7} & \textbf{82.86 } & 112.16  & \textbf{0.02 } & 69471.02  \\
    \bottomrule
    \end{longtable}
}

{
    \setlength{\tabcolsep}{0.6mm}
    \captionsetup[longtable]{font=normalsize}
    \renewcommand{\arraystretch}{0.8}
    \scriptsize   
    \begin{longtable}{ccc|ccc|cc|cc}
     \caption{
The comparison of different bounds on 15 large real-world graphs for $f(i)=\binom{i}{2}-s$ where $s=1,3$.}
    \label{tab:detailed-ub-kdc} \\
    \toprule
    \multirow{2}[4]{*}{\textbf{Graph}} & \multirow{2}[4]{*}{\textbf{s}} & \multirow{2}[4]{*}{\textbf{opt}} & \multirow{2}[4]{*}{\textbf{LP}} & \multirow{2}[4]{*}{\textbf{Simple}} & \multirow{2}[4]{*}{\textbf{SortBound}} & \multicolumn{2}{c|}{\textbf{Time(s)}} & \multicolumn{2}{c}{\textbf{Nodes($10^3$)}} \\
\cline{7-8} \cline{9-10}          &       &       &       &       &       & \textbf{\makecell{TwoDeg-\\BnB}} & \textbf{\makecell{TwoDeg-\\BnB/ub}} & \textbf{\makecell{TwoDeg-\\BnB}} & \textbf{\makecell{TwoDeg-\\BnB/ub}} \\
    \hline
    scc\_infect-dublin & 1     & 84    & 93    & 220   & \textbf{85} & \textbf{0.83 } & 550.84  & \textbf{0.01 } & 37268.88  \\
    $d_G=$83, $td_G=$219 & 3     & 84    & 93    & 220   & \textbf{87} & \textbf{1.14 } & -   & \textbf{0.54 } & - \\
    \hline
    web-indochina-2004 & 1     & 50    & \textbf{50} & 200   & 51    & \textbf{0.18 } & 1.08  & \textbf{0.00 } & 130.97  \\
    $d_G=$49, $td_G=$199 & 3     & 50    & \textbf{50} & 200   & 51    & \textbf{0.18 } & 63.23  & \textbf{0.00 } & 10206.45  \\
    \hline
    web-BerkStan & 1     & 29    & \textbf{29} & 60    & 30    & \textbf{0.03 } & 0.06  & \textbf{0.07 } & 30.64  \\
    $d_G=$28, $td_G=$59 & 3     & 29    & \textbf{29} & 60    & 32    & \textbf{0.05 } & 0.62  & \textbf{0.02} & 301.20  \\
    \hline
    web-webbase-2001 & 1     & 33    & \textbf{33} & 1680  & 34    & \textbf{0.33 } & 36.35  & \textbf{0.02 } & 2108.00  \\
    $d_G=$32, $td_G=$1679 & 3     & 33    & \textbf{33} & 1680  & 36    & \textbf{0.25 } & -   & \textbf{0.02 } & - \\
    \hline
    scc\_retweet-crawl & 1     & 21    & 26    & 196   & \textbf{23} & \textbf{0.09 } & 0.87  & \textbf{0.03 } & 175.32  \\
    $d_G=$19, $td_G=$195 & 3     & 21    & 26    & 196   & \textbf{25} & \textbf{0.16 } & 40.22  & \textbf{0.12 } & 10737.58  \\
    \hline
    ca-CondMat & 1     & 26    & \textbf{27} & 280   & \textbf{27} & \textbf{1.00 } & 7.11  & \textbf{0.05 } & 1370.79  \\
    $d_G=$25, $td_G=$279 & 3     & 26    & \textbf{27} & 280   & 29    & \textbf{1.00 } & 333.41  & \textbf{0.11 } & 72300.23  \\
    \hline
    tech-p2p-gnutella & 1     & 5     & 7     & 96    & \textbf{6} & \textbf{0.70 } & 16.62  & \textbf{0.01 } & 2213.99  \\
    $d_G=$6, $td_G=$95 & 3     & 5     & \textbf{8} & 96    & \textbf{8} & \textbf{0.72 } & 18.48  & \textbf{0.20 } & 27631.24  \\
    \hline
    rec-amazon & 1     & 6     & \textbf{6} & 9     & \textbf{6} & 0.08  & \textbf{0.16 } & \textbf{0.00 } & 5.85  \\
    $d_G=$4, $td_G=$8 & 3     & 6     & \textbf{6} & 9     & 8     & 0.08  & \textbf{0.16 } & \textbf{0.01 } & 2.90  \\
    \hline
    web-sk-2005 & 1     & 82    & \textbf{83} & 591   & \textbf{83} & \textbf{0.80 } & 7.71  & \textbf{0.09 } & 738.68  \\
    $d_G=$81, $td_G=$590 & 3     & 83    & \textbf{83} & 591   & 85    & \textbf{0.84 } & 1709.91  & \textbf{0.60 } & 262702.51  \\
    \hline
    web-uk-2005 & 1     & 500   & -   & 851   & \textbf{501} & \textbf{391.19 } & 576.81  & \textbf{0.19 } & 68.62  \\
    $d_G=$499, $td_G=$850 & 3     & 500   & -   & 851   & \textbf{501} & \textbf{365.31 } & 1861.92  & \textbf{2.24 } & 8681.00  \\
    \hline
    web-arabic-2005 & 1     & 102   & \textbf{102} & 1103  & 103   & \textbf{2.79} & 20.11 & \textbf{0.15} & 1258.14 \\
    $d_G=$101, $td_G=$1102 & 3     & 102   & \textbf{102} & 1103  & 105   & \textbf{2.91} & -   & \textbf{0.94} & - \\
    \hline
    ca-MathSciNet & 1     & 25    & -   & 497   & \textbf{26} & \textbf{7.92 } & 39.81  & \textbf{0.02 } & 6692.81  \\
    $d_G=$24, $td_G=$496 & 3     & 25    & -   & 497   & \textbf{28} & \textbf{7.62 } & 1228.93  & \textbf{0.08 } & 256325.02  \\
    \hline
    inf-roadNet-PA & 1     & 4     & -   & 10    & \textbf{5} & \textbf{1.82 } & 3.65  & \textbf{0.00 } & 1473.67  \\
    $d_G=$3, $td_G=$9 & 3     & 5     & -   & 10    & \textbf{7} & \textbf{1.54 } & 2.93  & \textbf{0.29 } & 1221.98  \\
    \hline
    inf-roadNet-CA & 1     & 4     & -   & 13    & \textbf{6} & \textbf{3.40 } & 5.59  & \textbf{0.01 } & 2544.21  \\
    $d_G=$3, $td_G=$12 & 3     & 5     & -   & 13    & \textbf{7} & \textbf{2.90 } & 4.57  & \textbf{0.19 } & 2073.22  \\
    \hline
    inf-road-usa & 1     & 4     & -   & 10    & \textbf{5} & \textbf{44.55 } & 98.89  & \textbf{0.01 } & 13238.70  \\
    $d_G=$3, $td_G=$9 & 3     & 5     & -   & 10    & \textbf{7} & \textbf{43.82 } & 88.33  & \textbf{0.11 } & 6874.49  \\
    \bottomrule
    \end{longtable}
}

\subsection{Relations Between Graph Size and Running Times}

\begin{figure}[htbp]
	\centering
    \scalebox{0.9}{
	\subfigure[$\gamma=0.99$]{
		\begin{minipage}[b]{0.45\textwidth}
			\includegraphics[width=1\textwidth]{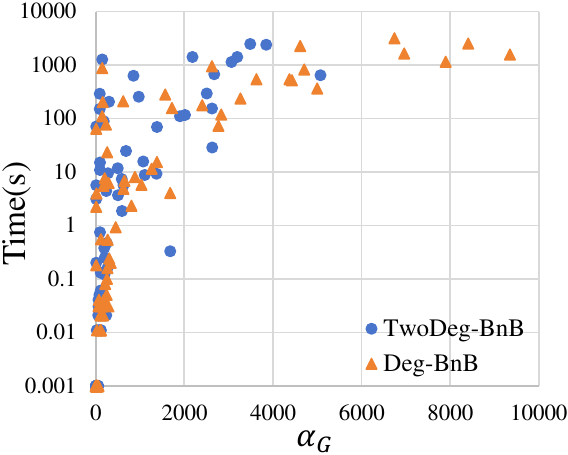} 
		\end{minipage}
		\label{subfig:d-0.99}
	}
        \subfigure[$\gamma=0.95$]{
            \begin{minipage}[b]{0.45\textwidth}
            \includegraphics[width=1\textwidth]{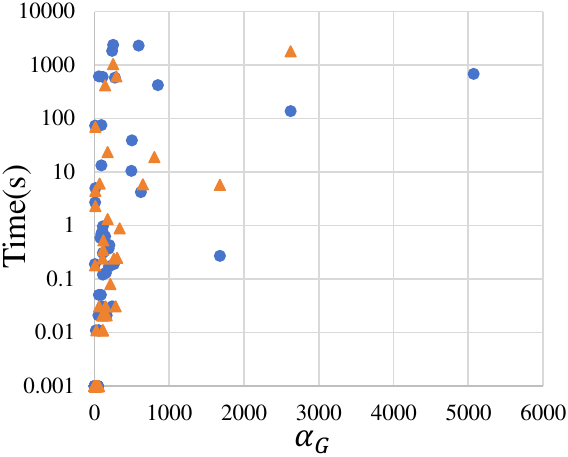}
            \end{minipage}
        \label{subfig:d-0.95}
        }
    }
	\caption{The relationship between the maximum subgraph size and the running time for $f(i)=\gamma\binom{i}{2}$ with $\gamma=0.99,0.95$ for algorithms TwoDeg-BnB and Deg-BnB.}
	\label{fig:d-t-cmp-qc}
\end{figure}

\begin{figure}[htbp]
	\centering
     \scalebox{0.9}{
    	\subfigure[$s=1$]{
    		\begin{minipage}[b]{0.45\textwidth}
    			\includegraphics[width=1\textwidth]{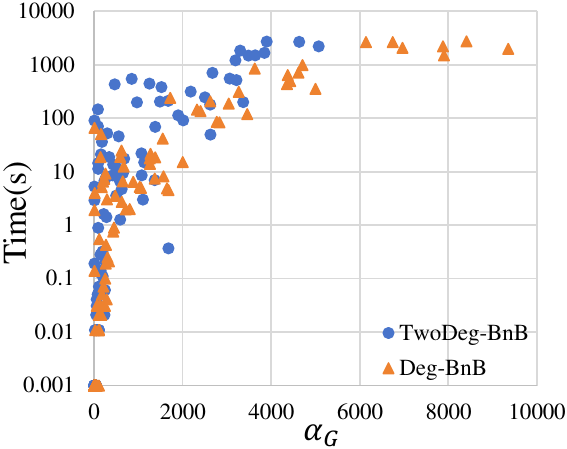} 
    		\end{minipage}
    		\label{subfig:kdc-d-k1}
    	}
            \subfigure[$s=3$]{
                \begin{minipage}[b]{0.45\textwidth}
                \includegraphics[width=1\textwidth]{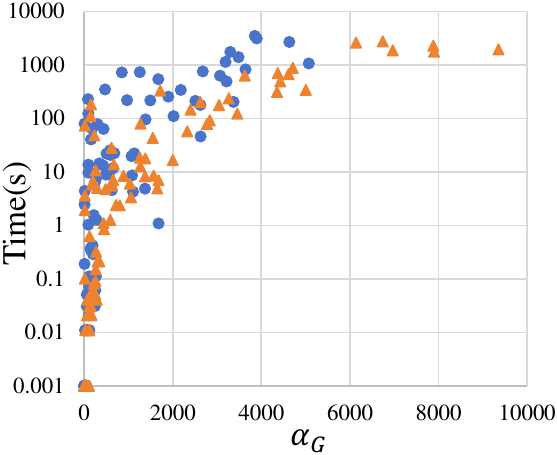}
                \end{minipage}
            \label{subfig:kdc-d-k3}
            }
        }
	\caption{The relationship between the maximum subgraph size and the running time for $f(i)=\binom{i}{2}-s$ with $s=1,\ 3$ for algorithms TwoDeg-BnB and Deg-BnB. }
	\label{fig:d-t-cmp-kdc}
\end{figure}

Lastly, we examine the relationship between the empirical wall-clock time of solving an instance and the \( \alpha_G \) value, i.e., the maximum size of the subgraphs obtained during graph decomposition. 
In Figures \ref{fig:d-t-cmp-qc} and \ref{fig:d-t-cmp-kdc}, each point represents an instance, which corresponds to a graph and a specific \( f(\cdot) \)-dense function. 
In nearly half of the solved cases, the largest subgraph contains at most 1,000 vertices, which is significantly smaller than the initial number of vertices. Generally, the running time shows a positive correlation with the maximum size of the subgraphs, supporting the rationale for minimizing \( \alpha_G \).



\section{Conclusion}

In this work, we introduced the concept of an \(f(\cdot)\)-dense graph, which generalizes various cohesive graph models (e.g., clique, \(\gamma\)-quasi clique, \(s\)-defective clique, and average $s$-plexes) by requiring at least \(f(n)\) edges for a vertex set of size \(n\). Building on the \(f(\cdot)\)-dense graphs, we studied the Maximum Low-Diameter \(f(\cdot)\)-Dense Subgraph (M\(f\)DS) problem, which seeks the largest subgraph constrained by both diameter at most two and the \(f(\cdot)\)-dense condition.

To solve M\(f\)DS, we studied its mixed integer linear program for M\(f\)DS. Moreover, we  proposed a practical exact algorithmic framework featuring a decomposition phase that decomposes the original problem into \(n\) smaller subproblems, as well as a branch-and-bound method which can be integrated into this framework. 
We provided a complexity analysis about the efficiency of these methods.
Extensive experiments on 139 real-world graphs and two \(f(\cdot)\)-dense functions demonstrate that the decomposition-based branch-and-bound approach achieves superior performance compared to other solution methods, including a standalone branch-and-bound or MIP solver. 

This work can be further extended from several perspectives. 
To the best of our knowledge, there are limited studies concentrating on the generalized clique model, e.g. \citep{veremyev2016exact}. 
In contrast, over the past decades, numerous investigations have been made for extracting dense subgraphs of specific forms, like robust vertex-connected subgraphs \citep{zhou2022effective,veremyev2014finding} or high-degree subgraphs \citep{balasundaram2011clique}. 
Therefore, it is still important to explore which graph structures hold significant importance  and which models could be further generalized in practical applications.

Another future research line of the work is the extension and parallelization of the graph decomposition framework. The graph decomposition only relies on the diameter-two bounded constraint; hence, we can easily extend it to other dense subgraph search algorithms.
Moreover, it is evident that once the graph is decomposed, calculations on the subgraphs do not depend on each other. Therefore, it is practical to implement parallel computing techniques to enhance the scalability of the proposed algorithm. As far as we know, the work in ~\citet{wang2022listing} has successfully scaled the large $k$-plexes extraction to huge graphs of 100 million vertices. 

\section*{Acknowledgments}
This work was supported by the National Natural Science Foundation of China under grant no. 62372093, and the Shenzhen Agency of Innovation under grant KJZD20240903095712016.






\bibliographystyle{apalike} 
\bibliography{cas-refs}

\appendix
\section{Missing Proofs}
\label{sec_missing_proof}
\subsection{Proof to Lemma 1}
\begin{proof}
    Suppose that $G=(V,E)$ is an $f(\cdot)$-dense graph. 
    By definition, $|E|\ge f(|V|)$, or  equivalently $\binom{|V|}{2}-|E|\le \binom{|V|}{2}-f(|V|)$. 
    Denote $\overline{E}$ as the edge set of $\overline{G}$, the complement graph of $G$.
    Then,$|\overline{E}|\le g_f(|V|)$.
    Because $g_f(\cdot)$ is monotonically non-increasing, for any induced subgraph of $\overline{G}$, say $\overline{H}=(V',\overline{E'})$, the inequality $|\overline{E'}|\le |\overline{E}|\le g_f(|V|)\le  g_f(|V'|)$ holds.
    This is equivalent to saying that the complete graph of $\overline{H}$, say $H=(V',E')$, satisfies $|E'|\ge f(|V'|)$, which completes the proof.
    \end{proof}
    
\subsection{Proof of Lemma 2}
\begin{proof}
        Because $f(\cdot)$ is positive and non-decreasing, there exists an integer $n_0\ge 2$ such that $f(n_0)>0$. Denoting $c_0=\binom{n_0}{2}$, then there exists $n_1>n_0$ such that $g_{f}(n_1)\geq c_0 \ge g_f(n_0)$.  The first inequality holds because $g_f(\cdot)$ is unbounded and the second holds because $g_f(\cdot)$ is non-decreasing.
        So, if we can construct an $f(\cdot)$-dense graph $G=(V,E)$ with $n_1$ vertices and find  an induced subgraph of $G$ which is not an $f(\cdot)$-dense graph, then we can complete the proof. 
        In other words, we now build a graph $\overline{G}=(V,\overline{E})$ of $n_1$ vertices such that $|\overline{E}|\le g_f(|V|)$ and show that the complement graph of one of the  induced subgraph of $\overline{G}$ is not an $f(\cdot)$-dense graph.
        Let us denote the vertex set $V$ as $\{v_1,..v_{n_0},\dots,v_{n_1}\}$. 
        To build the graph $\overline{G}=(V,\overline{E})$, we add edges that makes vertices $v_1,\dots,v_{n_0}$ as a clique and leaves the set of $v_{n_0+1},\dots,v_{n_1}$ as a independent set. 
        Then $|\overline{E}|=\binom{n_0}{2}\le g_f(n_1)$.
        So $\overline{G}$ satisfies $|\overline{E}|\le g_f(n_1)$.  That is to say, the complete graph of $\overline{G}$ is an $f(\cdot)$-dense graph by the definition of $f(\cdot)$-dense graph.
        
        Denote the subgraph induced by $V'=\{v_1,\dots.,v_{n_0}\}$ as $\overline{H}=(V',\overline{E'})$.  $\overline{H}$ is clearly a complete graph. 
        Because $g_f(n_0)=\binom{n_0}{2}-f(n_0)<\binom{n_0}{2}$ (note that $f(n_0)>0$), $|\overline{E'}|>g_f(n_0)$.
        So the complement graph of $\overline{H}$, denoted as $H$, is not $f(\cdot)$-dense graph. 
        
        In brief, we find that $G$ is an $f(\cdot)$-dense graph, but one of its induced graphs $H$, is not. Therefore, $f(\cdot)$ is not a hereditary-induced function.
    \end{proof}

\subsection{Proof of Lemma 3}
\begin{proof}    
    Suppose $S^*$ is an arbitrary maximum low-diameter $f(\cdot)$-dense subgraph in $G$.
    Denote that the first vertex in $S^*$ according to order $\pi$ is $v_i$. 
    Then it is clear that $S^*\setminus\{v_i\}\subseteq N_G^{(2)}(v_i) \cap  \{v_{i+1},\dots,v_n\}$. 
    Therefore, the $S^*$ is a maximum low-diameter $f(\cdot)$-dense subgraph in $G\left[\{v_i\}\cup \left\{ N_G^{(2)}(v_i) \cap  \{v_{i+1},\dots,v_n\}\right\}\right]$ which includes $v_i$.
    So, the largest maximum low-diameter $f(\cdot)$-dense subgraph over all $v_i$s is equal to $\omega_f(G)$.
\end{proof}

\subsection{Proof of Theorem 1}
\begin{proof}
    The time complexity of $T_{order}(G)$ is $O(|V|+|E|)$ as we explained.
   By definition, for any $v_i$, $N^{(2)}_{G_i}(v_i)= N_G(v_i)\cup \left\{N^{(2)}_G(v_i)\setminus N_G(v_i)\right\}$.
   Because $|N_G(v_i)\cap \{v_{i+1},\dots,v_n\}|\le d_G$ and $N^{(2)}_G(v_i)\setminus N_G(v_i)\le d_G\Delta_G$, then $\alpha_G\le d_G\Delta_G$.
    On the other hand,  $G_i$ is a subgraph of $G$ so $\alpha_G\le |V|$.
    Therefore, $\alpha_{G}\le \min (d_G\Delta_G, |V|)$.
\end{proof}

\subsection{Proof of Theorem 2}
\begin{proof}
    The bound of $T_{order}$ is a direct consequence of the two-hop degeneracy ordering.
    $\alpha_G \le td_G$ is simply because of the definition of $td_G$.
\end{proof}

\subsection{Proof of Lemma 4}
\begin{proof}
    Assume that for any maximum $f(\cdot)$-dense subgraph $S^*$ that $S\subseteq S^*\subseteq S\cup C$, $u\notin S^*$. 
\begin{itemize}
    \item Case 1: If $|N_G(u)|=|V|-1$, then it is clear that $S^*\cup \{u\}$ is still a low diameter $f(\cdot)$-dense subgraph, violating the assumption that $S^*$ is maximum.    
    \item Case 2: If $|N_G(u)|=|V|-2$, then there is only one vertex $v\in V$ such that $v\notin N_G(u)$. We separately discuss cases that $v\in S$ and $v\in S^*\setminus S$ in the followings.
            \begin{enumerate}
                \item If $v\in S^{*}\setminus S$, then $|E(S^*\cup \{u\}\setminus \{v\})|\geq |E(S^*)|\geq f(|S^*|)=f(|S^*\cup \{u\}\setminus \{v\}|)$. On the other hand,  the diameter of $G[S^*\cup \{u\}\setminus \{v\}]$ is bounded by 2 because every vertex in this subgraph is a neighbor of $u$. Therefore, $S^*\cup \{u\}\setminus \{v\}$ is an low-diameter $f(\cdot)$-dense subgraph contains $u$, conflict the assumption.                
                \item Now we consider the case $v\in S$. Firstly, we claim that there is a vertex $w$ in $S^*\setminus S$ such that at least one vertex in $S^*$ is not a neighbor of $w$, i.e., $S^* \nsubseteq (N_G(w)\cup \{w\})$.
                Suppose that for any $w'\in S^*\setminus S$, all vertices of $S^*$ are neighbors of $w'$.
               We denote $S^*\setminus S$ as $\{v_1,v_2,\dots,v_{|S^*\setminus S|}\}$. Then because $S\cup \{u\}$ is a low diameter $f(\cdot)$-dense subgraph, we can prove that $S\cup \{u\}\cup \{v_1\}$, $S\cup \{u\}\cup \{v_1,v_2\}$, \dots, $S\cup \{u\}\cup (S^*\setminus S)$ is all low diameter $f(\cdot)$-dense subgraphs by the same argument in Case 1. Therefore, we get a low diameter $f(\cdot)$-dense subgraph of size larger than $S^*$, which is a contradiction. So there must be a vertex $w$ in $S^*\setminus S$ such that one vertex in $S^*$ is not a neighbor of $w$. 
               
               Secondly, we claim that when there exists a $w\in S^*\setminus S$ such that $v$ is not the neighbor of $w$,  $G[S^*\cup \{u\}\setminus \{w\}]$ is also a low-diameter $f(\cdot)$-dense subgraph with the same size. 
               Clearly, $G[S^*\cup \{u\}\setminus \{w\}]$ is an $f(\cdot)$-dense subgraph.
               Now, we futher show that $G[S^*\cup \{u\}\setminus \{w\}]$ is a low diameter subgraph. Considering that all vertices in $S^*\setminus S$ are the neighbors of $v$,  we should prove that for any vertex pair $x$ and $y$ in $S^*\cup \{u\}\setminus \{w\}$, there exists a path connecting $x$ and $y$ of length at most 2. 
               \begin{enumerate}
                   \item If $x\neq v$ and $y\neq v$, then because $u$ is the neighbor of $x$ and $u$ is the neighbor of $y$, there must exist a path connecting $x$ and $y$ of length at most 2.
                   \item If $x= v$ and $y\in S$, because $diam(G[S\cup \{u\}])\leq 2$ we have there must exist a path connecting $x$ and $y$ of length smaller that or equal to 2.
                   \item If $x= v$ and $y\in (S^*\setminus S)\setminus \{w\}$, we discuss this in two situations. If $y= u$, then for any vertex $w'\in S^*\setminus S$ and $w'\neq w$, $w'$ is the neighbor of $v$ and $u$. Otherwise $y\neq u$, because $v$ is the neighbor of all vertices of $S^*\setminus S$, $x$ is the neighbor of $y$. Therefore, there must exist a path connecting $x$ and $y$ of length at most 2.
               \end{enumerate}
               In sum, $G[S^*\cup \{u\}\setminus \{w\}]$ is a low diameter $f(\cdot)$-dense subgraph with the same size as $G[S^*]$, which is contradicted to the assumption.

            \end{enumerate}
            
\end{itemize}
\end{proof}

\subsection{Proof of Lemma 5}
\begin{proof}
     If $S$ is not an $f(\cdot)$-dense subgraph, and $f(\cdot)$ is a hereditary-induced function, then any superset of $S$ is not an $f(\cdot)$-dense subgraph, so is the optimal solution $S^*$.
    
    If $|E(S\cup \{v\})|< f(|S|+1)$, then $G[S\cup \{v\}]$ is not an $f(\cdot)$-dense subgraph. The hereditary property indicates that any superset of  $S\cup \{v\}$ is not an $f(\cdot)$-dense graph.

\end{proof}

\subsection{Proof of Lemma 6}
\begin{proof}
     Denote $S'=S^* \setminus S$. So $S'\subseteq C$. Because $S^*$ is an $f(\cdot)$-dense subgraph, $|\overline{E(S^*)}|=|\overline{E(S\cup S')}|\le g_f(|S|+|S'|)$.
     Furthermore,  $\sum_{v\in S'}{|S\setminus N(v)|}+\overline{E(S)}$ is an underestimation of $|E(S^*)|$, i.e. $\sum_{v\in S'}{|S\setminus N(v)|}+|\overline{E(S)}| \le |\overline{E(S\cup S')}$|.
     That means $S'$ must satisfy $\sum_{v\in S}{|S\setminus N(v)|}+|\overline{E(S)}| \le g_f(|S|+|S'|)$.
     The computation of $k$ indicates that $S_k$ is the maximum subset in $C$ such that $\sum_{v\in S_k}{|S\setminus N(v_k)|}+\overline{E(S)}\le g_f(|S|+k)$.
     That is to say, $k$ is the upper bound of $|S'|$.
     This completes the proof.
\end{proof}

\subsection{Proof of Theorem 3}
\begin{proof}
   Denote $S'=S^*\setminus S$. 
   For each $i\in\{1,\dots,\chi\}$, denote $S'_i=S'\cap \Pi_i$ and $s'_i=|S'_i|$. 
   Then $\{S'_1,\dots,S'_\chi\}$ is a partition of  $S'$. Because each $S'_i$ is an independent set, $|\overline{E(S')}|\ge \sum_{i=1}^{\chi}\binom{s'_i}{2}$ holds.
   
   Furthermore, assume that each $\Pi_i$ is partitioned into $\sigma_i$ subsets by lines 7-13, i.e.  \(\Pi_i=\{R_{i}^1,R_{i}^2,\dots,R_{i}^{\sigma_{i}}\}\).   
   Denote $S'_{ij}=S'_i\cap R_i^j$ and $s'_{ij}=|S'_{ij}|$. Then \(\{S'_{i1},S'_{i2},\dots,S'_{i\sigma_i}\}\) is a partition of \(S'_i\).
   Clearly, $S'_{ij}$ is an independent set and the distance of every two vertices in $S'_{ij}$ is strictly larger than 2.
   
   Now, we have
   \begin{align}  
      & |\overline{E(S^*)}|=|\overline{E(S\cup S')}|=|\overline{E(S)}|+\sum_{v\in S'}{|S\setminus N(v)|}+|\overline{E(S')}| \nonumber\\
        &\ge |\overline{E(S)}|+\sum_{v\in S'}{w_S(v)}+\sum_{i=1}^{\chi}\binom{s'_i}{2}  \nonumber \\
        &= |\overline{E(S)}|+\sum_{i=1}^{\chi}\left(\sum_{v\in S'_i}{w_S(v)} + \binom{s'_i}{2}\right) \nonumber \\
        &= |\overline{E(S)}|+\sum_{i=1}^{\chi}\left(\sum_{j=1}^{\sigma_i}\sum_{v\in S'_{ij}}{w_S(v)} + \binom{s'_i}{2}\right).  \label{ineq_bound_1}
   \end{align}
   In the first inequality holds because  $w_S(v)$ is a substitution of  $|S\setminus N(v)|$  by definition and $|\overline{E(S')}|\ge \sum_{i=1}^{\chi}\binom{s'_i}{2}$.

   Note that $S^*$ is a two-hop constrained and there is at most one vertex in $S'_{ij}$ that is in \(S^*\). Let us denote \(\{R_{i}^{j_1},\dots,R_{i}^{j_k},\dots,R_{i}^{j_{s'_i}}\}\) as the $s'_i$ subsets that \(R_{i}^{j_k}\cap S'_i\neq \emptyset\).
    We further denote \(v_{m}^{(ij)}\) as the vertex in \(S'_{ij}\) with the minimum \(w_S(\cdot)\).  
  
    As indicated in line 5 in the algorithm \ref{alg:Sort-UB}, vertices in $\Pi_i$ are ordered by $w_S(\cdot)$ in non-increasing ordering. We assume the order of vertices in $\Pi_i$ as $\langle v_i^1,v_i^2,\dots,v_i^{|\Pi_i|} \rangle$. 
    By line 9 of alg. \ref{alg:Sort-UB},  we have \(w_S(v_{m}^{(i1)}) \le \dots \le w_S(v_{m}^{(ij)})\le \dots \le w_S(v_{m}^{(i\sigma_{i})})\).

Therefore, for every $i\in\{1,\dots,\chi\}$
  \begin{align} 
    &\sum_{j=1}^{\sigma_i}\sum_{v\in S'_{ij}}{w_S(v)} + \binom{s'_i}{2} \nonumber \\  
    & \ge \sum_{j=1, S'_{ij}\cap R_i^j\neq \emptyset}^{\sigma_i}w_S(v_{m}^{(ij)}) + \binom{s'_i}{2} \nonumber\\
    & =\sum_{k=1}^{s'_i}w_S(v_{m}^{(ij_k)}) + \binom{s'_i}{2}\nonumber\\
    &\ge \sum_{j=1}^{s'_i}\left(w_S(v_{m}^{(ij)})+j-1 \right) = \sum_{j=1}^{s'_i}w'_S(v_{m}^{(ij)}).\label{ineq_bound_3}      
    \end{align}
    
The first inequality holds because at most one vertex in $S'_{ij}$ can be added into \(S^*\). The second inequality holds because of the ordering .
Combining Inequalities (\ref{ineq_bound_1}) and (\ref{ineq_bound_3}), we further obtain   
      $|\overline{E(S^*)}|\ge |\overline{E(S)}|+\sum_{i=1}^{\chi}\sum_{j=1}^{s'_i}w'_S(v_{m}^{(ij)}).$

   
   Furthermore, in line 18 of in the algorithm \ref{alg:Sort-UB}, the vertices in $C'$ are ordered such that $w'_S(v_1)\leq w'_S(v_2)\leq,\ldots,\leq w'_S(v_{|C'|})$. So we have 
   \[
   |\overline{E(S^*)}|\ge |\overline{E(S)}|+\sum_{i=1}^{\chi}\sum_{j=1}^{s'_i}w'_S(v_{m}^{(ij)})  
   \ge |\overline{E(S)}|+ \sum_{j=1}^{|S'|}w'_S(v_j).
   \]
   Therefore, $|\overline{E(S)}|+\sum_{j=1}^{|S'|}w'_S(v_j)$ is an underestimation of $|\overline{E(S^*)}|$.
   Following the argument in Lemma 6, the largest $k$ obtained by lines 19-20 in this algorithm is the upper bound of $|S'|$. That is to say $|S|+k$ is an upper bound of $|S^*|$.
\end{proof}

\end{document}